\documentclass[iop]{emulateapj}
\usepackage[breaklinks,colorlinks,citecolor=blue,linkcolor=magenta,urlcolor=blue]{hyperref}
\usepackage{subfigure}

\newcommand*{\mysub}[2]{\ensuremath{#1_{\mathrm{#2}}}}

\newcommand*{\Omegam}{\mysub{\Omega}{m}}

\newcommand*{\Omegal}{\mysub{\Omega}{\Lambda}}

\newcommand*{\LCDM}{$\Lambda$CDM }

\newcommand*{\Msun}{\ensuremath{\, M_{\odot}}}

\newcommand*{\ltsim}{\ {\raise-.75ex\hbox{$\buildrel<\over\sim$}}\ }
\newcommand*{\gtsim}{\ {\raise-.75ex\hbox{$\buildrel>\over\sim$}}\ }
\newcommand*{\proptosim}{\ {\raise-.75ex\hbox{$\buildrel\propto\over\sim$}}\ }

\newcommand{\Spitzer}{{\it Spitzer }}
\newcommand{\HST}{{\it Hubble Space Telescope }}

\shorttitle{Spitzer's View of the CCPC}
\shortauthors{Franck \& McGaugh}

\begin{document}
\title{Spitzer's View of the Candidate Cluster and Protocluster Catalog (CCPC)}


\author{J.R. Franck\altaffilmark{1},
	 S.S. McGaugh\altaffilmark{1}}
\altaffiltext{1}{Case Western Reserve University, 10900 Euclid Ave., Cleveland, OH 44106}

\begin{abstract}
The Candidate Cluster and Protocluster Catalog (CCPC) contains 218 galaxy overdensities composed of more than
2000 galaxies with spectroscopic redshifts spanning the first few Gyrs after the Big Bang $(2.0\le z<6.6)$. 
We use \Spitzer archival data to track the underlying stellar mass of these overdense regions 
in various temporal cross sections by building rest-frame near-infrared luminosity functions 
across the span of redshifts.This exercise maps the stellar growth
of protocluster galaxies, as halos in the densest environments should be the most massive 
from hierarchical accretion.  The characteristic apparent magnitude, $m^*(z)$, is relatively flat from $2.0\le z<6.6$,
consistent with a passive evolution of an old stellar population. This trend maps smoothly to lower redshift 
results of cluster galaxies from other works. We find no difference in the luminosity 
functions of galaxies in the field versus protoclusters at a given redshift, apart from their density. 
\end{abstract}


\section{Introduction} \label{sec:intro}

The nascent study of protoclusters is at a juncture of two 
important evolutionary epochs in the universe: the early growth 
of large structures and the rapid assembly of galaxy mass at $z\ge2$. Careful 
study of these objects can probe both cosmology and galaxy formation and evolution.
The initial mass overdensities in the early universe (and their subsequent collapse) are governed by the 
cosmic matter density (\Omegam), $\sigma_8$, and the cosmological constant \Omegal. 
Thus, by examining their properties (mass, evolutionary state) and number density of these early structures, they can 
provide constraints on these cosmological parameters. The tracers of these overdense structures (i.e., galaxies) 
can be used to investigate the role environment plays in their growth and evolution. Formation models of galaxies
must match observations across both cosmic time and throughout space to be considered viable. More simply, void
and cluster galaxies must both be reproduced for all $z\ge0$. The focus of this work will be primarily on the
properties of galaxies in dense environments, but this juncture of structure and galaxy evolution are clearly 
relatable in many ways.


It has been clear for many decades that galaxies at $z=0$ have varied properties that 
correlate strongly with density in the well-established morphology-density relation \citep{dre80,bal04}.
Galaxies in dense environments show clear evolution at higher redshift \citep{but84}, in that 
there are more passive galaxies in local clusters, but a greater fraction 
of star forming objects in higher redshift systems. This can be seen observationally in the fraction of quiescent, `red-sequence' objects 
in contrast with the `blue cloud' of star forming galaxies (SFGs) as a function of redshift \citep{sta95,rak95}.  These two facts provide 
an initial scaffolding from which galaxy evolutionary models can be
wrought: quiescent systems dominate high density regions of the local universe, and this need
not hold throughout cosmic time. What is the path that must be taken to satisfy these two
simple observations?

One prescription to turn these observational facts into a coherent model is to identify physical 
processes that effectively turn SFGs into quiescent systems preferentially
in dense environments. Commonly invoked interactions are ram-pressure stripping of cold gas
by the intracluster medium \cite[ICM;][]{gun72}, removal of the hot gas halo of a galaxy to halt the
cooling  of gas to sustain star formation \cite[e.g. strangulation;][]{lar80}, galaxy-galaxy interactions 
that disrupt the galaxy \citep{moo96}, and a variety of others. The overarching umbrella that is used
to refer to these proposed mechanisms is called `quenching' \citep{pen10}, loosely defined here as 
the environmental process(es) that abruptly discontinue star formation in a galaxy. 

Each proposed quenching process has a distinct time scale at which it can effectively halt star formation. These are generally related to the crossing/dynamical time
of the galaxies interacting with other galaxies and/or the ICM. For this latter case, it also assumes that the hot ICM
is in place at the epoch in question so as to produce the desired effect. 
These quenching timescales have been estimated to operate on the order of a few Gyrs for an effective change 
to manifest itself \citep{bro13}. If cluster galaxies are transformed primarily by one 
quenching mechanisms, it might be possible to examine the galaxy populations in clusters at various cross-sections 
in time (e.g., redshifts) to isolate the process responsible.

Another proposed scenario used to explain the evolution of the galaxy population in clusters
is from galaxy mergers. Hierarchical accretion suggests that galaxies assemble from the bottom 
up. In the context of clusters, these mergers have been proposed to be either dry \cite[e.g. gas-free;][]{van05} 
or wet \cite[gas-rich collisions;][]{fab07}. In the wet merger scenario, two or more blue cloud galaxies come together
in a burst of star formation, turning into a spheroidal configuration \citep{mih96}, and are then quenched via some mechanism. 
This is effectively a transition from the blue cloud onto the red sequence. Dry mergers, on the other hand, push galaxies
along the red sequence as they grow in stellar mass, but remain with a generally passive stellar population (color).
It is possible and even probable that the various quenching mechanisms and merger scenarios could all play
a role of varying importance. Do any observations hint at a timescale of rapid development in the history 
of cluster galaxies?

The red sequence feature of galaxy clusters has been historically used as a tracer of the core 
galaxy population at redshifts $z<1.5$ \citep{sta95,rak95,sta98,eis08,mei09}. Using models of galaxy colors
 a mean stellar age of the individual systems can be estimated.
Progenitor bias \citep{van01} can ultimately lead to an increased estimated formation redshift ($z_f$) for a cluster's 
galaxy population as a whole. Despite this caveat, the stellar age of the brightest systems on the red sequence 
can give some indication as to when the most massive galaxies formed their stars.  Models of galaxy growth used
to match the evolution of the red sequence feature typically rely on pure, passive evolution models that were 
formed in a single burst at high redshift $z_f>3$ \citep{eis08}. In fact, some studies of clusters suggest even 
larger formation redshifts of $z_f>5$ \citep{rak95,sch09}. As the redshift of the cluster sample increases, the formation 
redshift also grows (with an accompanying increase of scatter). 

For instance, for cluster redshifts $z>1$, it appears that
$z_f=30$ is not ruled out \cite[Fig 19 of][]{eis08}. Not 
all cluster galaxies need form at $z_f>5$, but at least some passive systems were born remarkably early in the universe.
This could simply be a manifestation of galaxy downsizing \citep{cow96}, in which the largest stellar mass galaxies (presumably formed
in the densest regions) formed earlier in the universe. Within the literature, there is no consensus for any single model or mechanism for the redshift evolution of the red sequence 
feature \citep{fas14,fri14}. 

By shifting the focus from the highly biased red sequence galaxies to cluster populations in general,
there is the hope that the bulk stellar properties of these systems could be investigated as a function
of redshift. \citet{man10} mapped the luminosity functions (LFs) of cluster galaxies spanning $0.3<z<2.0$ 
at \Spitzer wavelengths. At the highest redshift of this sample, \Spitzer 3.6 $\mu$m coverage measures 
rest-frame $J$ band, which is a tracer of the stellar population for a range of ages. They mapped the 
evolution of the characteristic luminosity $m^*(z)$\footnote{We designate the characteristic magnitude in lower case ($m^*$) to emphasize
that it is a measure of the apparent magnitude and to distinguish it from the stellar mass ($M_{\star}$) also found
in this text.} of the clusters by comparing the data
to models of simple stellar populations with various formation redshifts $z_f$. This is similar
to the exercise performed for red-sequence fitting. The mean formation age of these systems was $z_f \sim 2.5$ \citep{man10},
with the same behavior noted previously: higher redshift clusters favor higher formation redshifts. Their two 
highest redshift bins ($z\ge1.5$) have $m^*(z)$ values nearly a magnitude fainter than the predicted 
evolution of their best fitting model (their Fig 7). The conclusion drawn from this observation is that rapid mass assembly (up to $4\times$ growth)
must occur in cluster galaxies $z\le1.5$ \citep{man10}. \citet{bro13} investigated the star formation activity of 
galaxies in these clusters to look for clues as to the nature of this mass assembly. They found that the star 
formation within the core of these clusters transition from unquenched to quenched at the same
epoch ($z\sim1.4$) as the rapid assembly era within \citet{man10}. This behavior is generally 
attributed to wet mergers within the cluster core, rapidly growing the mass of these systems and 
then abruptly turning off the star formation activity. 

In the previous examples, all of the structures were considered to be clusters.
Generally speaking, the highest redshift at which virialized halos of $M\ge10^{14}$ \Msun (e.g. clusters)
are expected is at $z\sim 2$ in large \LCDM simulations \citep{chi13}. The collection of components that will constitute 
a cluster in the future is referred to as a protocluster. Observationally, galaxy overdensities at $z>2$ are designated 
as protoclusters for the sake of simplicity, as it is difficult to confirm these systems to be in virial equilibrium apart
from a handful of cases \citep{gob11,wan16}. This unique transition point in the universe represents an epoch at which galaxies
could first begin to interact with one another. \citet{man10} and \citet{bro13} presented tantalizing evidence that the majority of 
mass assembly occurred around $z\sim1.5$, but higher redshift luminosity functions of structure might yield further insight into the 
galaxy growth within dense environments. 

\citet{wyl13} identified \Spitzer galaxy overdensities around high redshift ($1.3<z<3.1$) radio-loud AGN and built
3.6 and 4.5 $\mu$m LFs in \citet{wyl14}. This redshift range overlapped the sample of \citet{man10} and extended the
age probed by more than 1 Gyr. Remarkably, the $m_{3.6}^*$ and $m_{4.5}^*$ evolution over the redshifts probed are 
well fit by a passive stellar evolution model formed at $z_f=3$ or larger. They also do not match the results of \citet{man10},
in that they fail to see a burst of mass assembly at $z\le1.5$. This is attributed to a sampling bias, in that the high redshift 
overdensities are thought to be the most massive, rare systems in the universe, while the lower redshift sample is tracing the 
growth of less massive clusters. This is analogous to galaxy downsizing, in that the most massive overdensities are fated to assemble 
into a cluster mass halo more quickly in \LCDM \citep{chi13}. Therefore, signatures of mass assembly for the progenitor systems of 
\citep{wyl14} could potentially be observable beyond their redshift limit of $z\approx3$. In Section~\ref{sec:results}, we probe these earlier epochs to possibly identify epochs
at which rapid mass growth or quenching might be exhibited. 

Thus far, it appears that galaxies within clusters, as traced by both red sequence and LF models, form 
at high redshift ($z_f\ge3$) and evolve passively thereafter. This is an interesting result, as the cosmic star formation 
rate in the universe does not peak until approximately $z\sim2$ \citep{mad14}. Indeed, if galaxies in dense environments 
form earlier than their `field' galaxy counterparts, which follow the mean trend, then evidence of this should be 
apparent at high redshifts. The number of spectroscopically confirmed protoclusters has evolved considerably after the first few
discoveries \citep{ste98,ven02}, but were still only numbered in the few dozens up until recently. These were also identified by a wide 
range of selection techniques, from blind spectroscopic surveys \citep{ste98} to targeted narrowband (NB) imaging around high redshift
quasars \citep{ven07}. In the instances in which these galaxy overdensities were compared with field galaxies at a similar redshift,
the results of environmental evolution are varied at best. We continue the exploration of these results at higher redshift in 
Section~\ref{sec:dis}.


The majority of cases in the literature where protocluster galaxies were measured with respect to field sources
consist of one or two candidate structures that are compared to a `blank' field-of-view. For instance,
\citet{cas15} studied a protocluster at $z=2.5$ within the COSMOS field and found that it had evidence of greater
AGN activity, more indications of merging/interacting galaxies, and a population of Lyman-Break Galaxies (LBGs) with
$\sim1.5\times$ greater stellar mass. These had similar star formation rates (SFRs) when compared to field sources, though.
For other protoclusters at $z\le2.5$ identified with NB filters centered on the redshifted $H\alpha$ line, the candidate galaxies were also found 
to be dustier \citep{coo14}, more massive, and not significantly forming more stars than their field counterparts \citep{hat11}. 

Similar studies 
that trace the $Ly\alpha$ emission of protocluster galaxies find that they are generally brighter \citep{zhe16}, \emph{less} dusty (although this may be a selection 
effect), and younger \citep{dey16}. $Ly\alpha$ equivalent widths (EWs) have been used extensively as an estimator for the star formation 
rate of galaxies in the high redshift universe \citep{2010MNRAS.401.2343D}, and \citet{zhe16} find evidence that the EWs are stronger at $z=2.8$. \citet{dey16} do
not find such EW dependence at $z=3.8$, while \citet{tos16} finds smaller EW$_0$ in a protocluster at $z=3.67$ when compared to the field. \citet{hay11} and \citet{hay12} find only the reddest galaxies in their structures have statistically
significant environmental dependence. It seems clear that when individual high redshift structures are analyzed, usually an environmental influence is found,
but the effect is varied. It is also apparent that in some protoclusters, the property in question is enhanced (e.g. dustier galaxies), while in others it is
diminished, even at similar redshifts.

This does not seem to be the case when multiple candidate structures are identified in the same manner, or large surveys are systematically 
analyzed. \citet{own16} used a constant number density selection technique for the UKIDSS survey to measure the evolution of galaxies. This
selection technique is thought to be much less-biased when compared with a mass-limited selection in matching progenitor galaxies to their offspring.
Their results point to a relatively early formation redshift ($z\ge3$ and possibly earlier) 
and subsequent passive evolution with little environmental influence. With a similar method, \citet{zha16} tracked the growth and evolution of $z=0$ Brightest Cluster Galaxies (BCGs) from $z\sim2$. A key 
result is that most of the mass growth must occur at $z<2$, as BCGs are not divergent from similar mass systems. \citet{die13,die15} identified more than 40 spectroscopic galaxy group 
and larger systems in the COSMOS field. Compared to field sources, their analysis revealed no statistically significant color (stellar population) difference
in the group environments with respect to the field.

This begs the question of why overdensities do not show up as significant deviations from their
field counterparts when analyzed systematically at a variety of redshifts, while individual systems have found statistically different evolution over similar epochs. 
It is often difficult to match the results of these protocluster studies coherently. These protoclusters exist at a wide range of redshifts, each within their
infancy and characterized by the rapid changes expected in a \LCDM universe \citep{chi13,mul15}. Particularly for the individual case studies \citep{hat11,hay12,coo15,dey16,zhe16},
different instruments and selection techniques were used, which targeted different populations of galaxies. Furthermore, the galaxy properties themselves were not all analyzed  
in the same manner.


In an attempt at tackling the complex problem of galaxy evolution, it can be helpful
to simplify the approach \citep{abr16}, and confront the issue in a new way instead of adding further epicycles.
It was this impetus that inspired us to construct the Candidate Cluster 
and Protocluster Catalog \cite[hereafter CCPC;][]{ccpc1,ccpc2}. With a straightforward algorithm, 
we were able to systematically detect galaxy overdensities from disparate spectroscopic catalogs 
in the high redshift universe. Then, the properties of the galaxies within 
these candidate structures can be traced through cosmic time in a series 
of cross-sections. We hope to address the evolution of galaxy stellar mass, 
as \citet{man10} and \citet{wyl14} did at lower redshifts, while simultaneously
mapping the field evolution in a consistent manner over a range of redshifts. Although not a longitudinal study, these
snapshots of galaxies in dense environments may provide a powerful  glimpse into the 
behavior governing their evolution. 

We present here a detailed analysis of the CCPC sample to date
using \Spitzer IRAC and supplementary \HST near-infrared (NIR). 
In this back-to-basics approach, we measure the 3.6 and 4.5 $\mu$m LFs of the galaxies in the CCPC as a function 
of redshift. This serves as a tracer of the stellar mass of these objects, and are compared with `field' galaxies
identified in the same spectroscopic surveys used for the CCPC. 

Throughout this work we assume a cosmology of $\Omegam = 0.3$ and $\Omegal = 0.7$, 
with a Hubble value of $H_0 = 70$ km s$^{-1}$  Mpc$^{-1}$. All magnitudes quoted 
are in the AB system, with apparent magnitudes
in the four \Spitzer IRAC channels denoted as [3.6], [4.5], [5.8], and [8]. 
The accompanying apparent magnitudes from \HST measurements will be referred to by the 
filter name (e.g. $F160W$).

\section{Observations} \label{sec:obs}

The CCPC identifies structure around galaxies by mining archival
spectroscopic redshift catalogs. Any volume within a search radius of
$R=20$ comoving Mpc (cMpc) and distance in redshift space $\Delta z$ 
corresponding to $\pm20$ cMpc which contains 4 or more galaxies and display a 
galaxy overdensity of $\delta_{gal}>0.25$ is considered a candidate system \citep{ccpc1,ccpc2}. These are 
the minimum requirements and, in many cases, are exceeded.

The algorithm was designed to be used on a variety of survey depths and widths, where 
$N\ge4$ is used as a signpost from which the volume overdensity can be computed. The average
number density of these systems is $n\sim0.05$ cMpc$^{-3}$, and a mean galaxy overdensity
of $\delta_{gal}\sim2.0$. We have shown in \citet{ccpc1} and \citet{ccpc2} that these protocluster candidates
are statistically distinct both spatially and along the line of sight from non-CCPC 
galaxies.  

\subsection{Galaxy Selection}

The catalog contains a total of 216 structures spanning $2<z<6.56$.  We include two objects 
of interest at $z=6.56$ from \citet{ccpc2} to bring the total number of candidate systems to 218. These systems lack 
field galaxies, and so a galaxy overdensity ($\delta_{gal}$) cannot be computed, a requirement for inclusion in the CCPC. In this list there 
exist 2048 galaxies. The vast majority of these objects are identified 
either by spectroscopically targeting Lyman Break Galaxies \cite[LBGs][]{ste98} or follow-up
spectroscopy on suspected $H\alpha/Ly\alpha$ emitters from NB imaging \citep{ven07}.

We note here briefly that SFGs are not the dominant 
massive galaxy population at these epochs. In the high redshift universe, \citet{van06} estimated
that only 20$\%$ of all galaxies with $M_{\star}>10^{11}$ $\Msun$ are LBGs. The majority of systems ($\sim70\%$) 
are thought to be Distant Red Galaxies (DRGs), which are often too faint in the observed optical passbands 
to be spectroscopically targeted. They can also lack the strong emission lines of their unobscured 
star-forming counterparts. We will discuss further implications of this in Section~\ref{sec:dis}.

\subsection{Data}

All of the measurements in this work came from archival data sets.
The \Spitzer Heritage Archive provided nearly 600 processed images in all four IRAC
channels covering most of the CCPC fields. The greatest wavelength coverage of these
systems came from the first two channels (3.6 and 4.5 $\mu$m), with galaxy contributions 
from 177 and 184 CCPC structures, respectively. Nearly $\sim75\%$ of the 
2048 CCPC galaxies were measured at 4.5 $\mu$m. All photometry was performed using 2'' radius apertures 
using the IRAF\footnote{IRAF is distributed by the National
Optical Astronomy Observatories, which are operated by the Association of
Universities for Research in Astronomy, Inc., under cooperative agreement with the
National Science Foundation.} \citep{iraf86,iraf93} QPHOT package, and the
magnitudes were computed by $m_{AB}=23.9 - 2.5 \log$($f_{\nu}/1 \mu$JY). Aperture corrections were applied to each galaxy 
(0.32 mags in [3.6] and 0.36 for [4.5]), as in \citet{pap10}.
Our photometry is consistent with the reported [3.6] and [4.5] AB magnitude values
compiled in the 3D-HST database \citep{ske14} within the CANDELS fields, apart from the 
aperture corrections.

We obtained more limited data coverage of the CCPC with the \HST in the
$F160W$ bandpass via the \emph{Hubble Legacy Archive}. The scope of these images were primarily concentrated within the 
CANDELS fields \citep{CANDELS}. Approximately $25\%$ of the CCPC was measured within these
images. Fluxes were 
measured in apertures of radius 0.4'' with a zeropoint magnitude of 25.75, values which were adopted from the 
WFC3 Handbook. The pixel scale of these images ranged from $\sim$7 to 33.33 pix/''.

With \HST and \Spitzer data, there were cases in which a galaxy was measured 
in a number of images. All magnitudes listed here are the uncertainty-weighted 
mean value of the photometry.

\subsection{Building the Luminosity Function}

We map the evolution of the CCPC galaxy luminosity function using the \citet{1976ApJ...203..297S}
form 
\begin{equation}
\phi(L)(d L) = \phi^* \big(\frac{L}{L^*}\big)^{\alpha}e^{-L/L^*} \frac{d L}{L^*},
\end{equation}
which relates the characteristic number density ($\phi(L)$) of sources over a range
of luminosities. $L^*$ is the characteristic luminosity of the distribution where the
number density decreases rapidly, and $\alpha$ is the slope of the faint end. The scaling 
factor of the Luminosity Function is $\phi^*$. In this work, we will adopt the magnitude functional
form of
\begin{equation} \label{eq:LF}
\phi(m) = 0.4 \ln(10) \phi^* \frac{10^{0.4(m^*-m)^{\alpha+1}}}{\exp{\big[10^{0.4(m^*-m)}\big]}}.
\end{equation}

To construct the distribution function, we first compute the density of each candidate 
structure by finding the minimum rectangular region (in units of arcmin$^{2}$) that bounds all galaxies. We then place 
galaxies in magnitude bins of $\Delta m=0.4$ for a given redshift range, and finally divide this by the summed surface area of candidates
at that redshift. The uncertainty for each magnitude bin is the 1$\sigma$ photometric uncertainty of the galaxy magnitudes 
in that bin. The number density uncertainty is Poissonian ($\sqrt{N}$).  The redshift
bins were designed to offer a balance between similar temporal spacing and number of galaxies 
at each epoch. This balance is necessary, for if the time period probed by each snapshot is too variable 
or too large, then any evolutionary inferences that the luminosity function might provide could be 
lost. In addition, too few galaxies in a redshift bin can be insufficient in fitting the parameters
of the Schechter function. We aim for $N\ge10^2$ galaxies per redshift bin whenever possible, as this
provides a robust fit to the data in practice.

We computed the $e-$ and $k-$ corrections of each galaxy 
by estimating \cite[using EzGal;][]{ezgal} the observed magnitude difference of a BaSTI simple stellar population model 
\citep{basti} between the model's magnitude at the observed redshift of the galaxy and the center of its redshift bin. Typically these
corrections are less than $\Delta m\sim0.05$ magnitudes, which are smaller than our photometric uncertainties. There is no
net change in the values of $m^*$ when the corrections are applied, as these shifts in magnitude are balanced out by the galaxies at either
end of the redshift bin.

For fitting the Schechter function to the data, we used SciPy's \emph{curvefit} routine. This routine takes 
the data table, the Schechter function, the parameters to be fit and a set of initial guesses
for those parameters. These initial input values are insensitive to the outcome.
 The solution to Equation~\ref{eq:LF} is optimized via a Levenberg-Marquardt algorithm.
In practice, if the data have sufficient depth to fit the faint-end slope $\alpha$, all three parameters can 
be solved for simultaneously, as \citet{wyl14} were able to do. However, the variability of our data depth does 
not permit us to reliably fit $\alpha$ at all redshifts, and we therefore set it as a constant $\alpha=-1$.  This 
is standard procedure for \Spitzer LFs of insufficient depth \citep{man10}. \citet{wyl14} obtained the same 
results independently of whether $\alpha$ was fixed or not. The selection function of spectroscopic surveys 
used to construct the CCPC will typically favor bright galaxies out of necessity, which would artificially 
restrict the faintest regions in magnitude space, regardless. 
   
The main science goal of these LFs is to investigate the temporal evolution of $m^*(z)$ and number density of the largest 
galaxies in these systems, and so the faint slope of the galaxies is of relatively minor importance. Even at low redshifts, 
determining $\alpha$ can be problematic, as low surface brightness galaxies are often missed \citep{mcg96}. When the value
of $\alpha$ is allowed to vary, it is generally consistent with $\alpha\approx-1$ within the uncertainties for the lowest 
redshift sources in our data. At higher redshifts, there are too few galaxies in the faint magnitude bins from incompleteness,
and the fitting routine breaks down. In short, the optimization procedure of \emph{curvefit} provides a more 
robust fit to the data with a constant $\alpha=-1$. There have been no completeness corrections implemented on the data set.
The focus of this research is on tracing the brightest, most massive galaxies at a given epoch. Attempting to adjust the number of 
faint galaxies is (1) not important in achieving the research aim and (2) uncertain at best, as spectroscopic surveys 
of high redshift sources are inherently biased in this regard. 

Once the fitting routine has provided values of $m*$ and $\phi^*$, the uncertainties are calculated via bootstrapping.
The galaxies in each redshift bin are resampled with replacement $10^4$ times. Each instance is fitted to Eq ~\ref{eq:LF}
in the same manner as the full data set. The 95$\%$ confidence region of the data is provided by fitting the $\pm2\sigma$ 
values of $m^*$. Figure ~\ref{fig:z3_LF} illustrates this.

\section{Results} \label{sec:results}

\subsection{Luminosity Functions}

\begin{figure}
\centering
\includegraphics[scale=0.5]{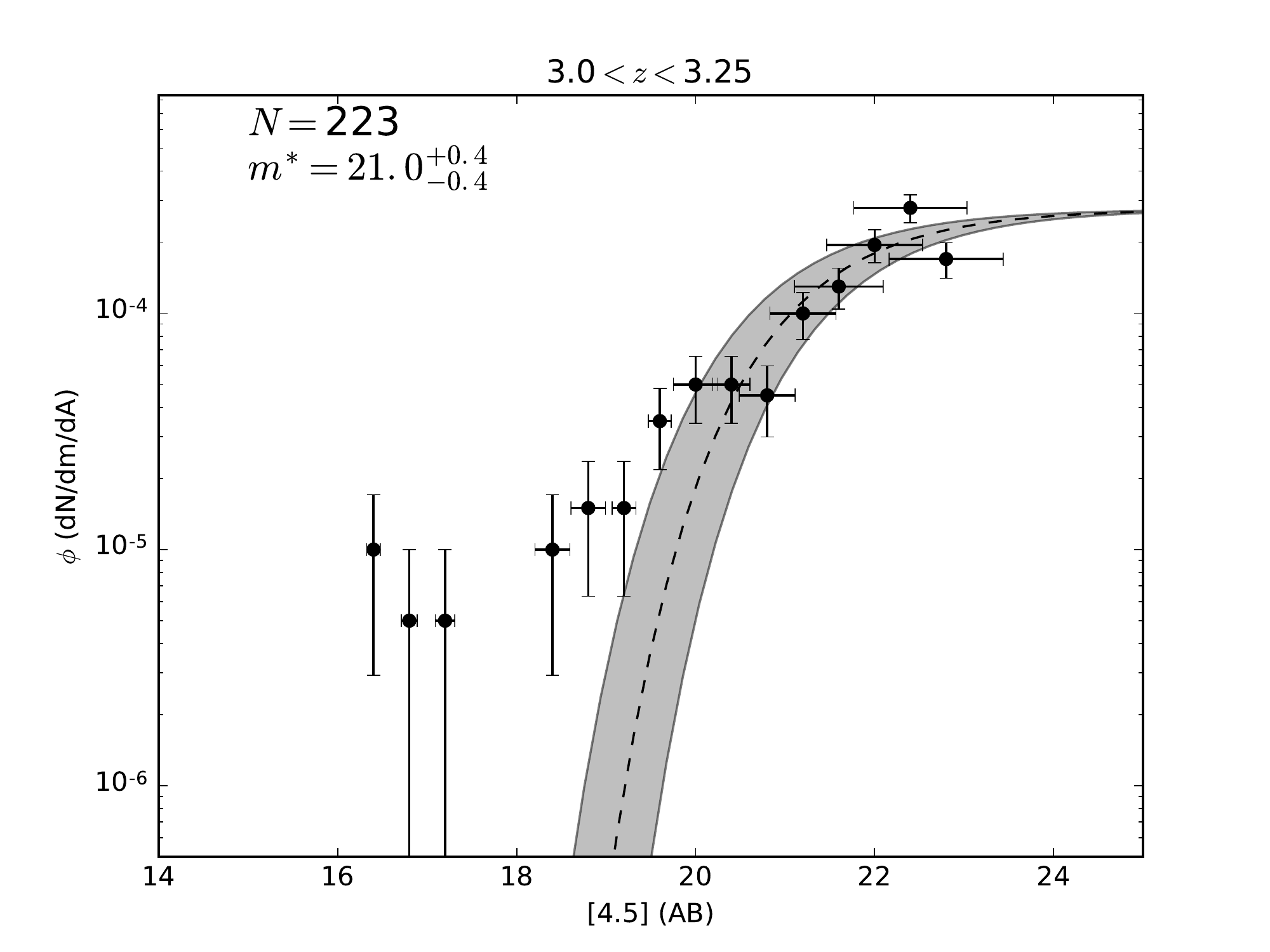}
\hfill
\caption{The 4.5$\mu$m Luminosity Function of CCPC galaxies at redshifts $3.0\le z<3.25$.
$\phi$ is the number of galaxies at a specific magnitude per square arcminute ($dN/dm/dA$). Uncertainties
in the number density of sources are Poissonian ($\sqrt{N(dm)}$), while the magnitude uncertainties in each bin
are the average uncertainty of the bin's galaxies. SciPy's \emph{curvefit} 
routine's best fit to the data is the black dashed line. The number of galaxies
in the plot and the value of $m^*$ and its $2\sigma$ bootstrapped uncertainties
are listed in the top left corner. In this work, $\alpha$ is defined 
to be -1. The gray 
shaded area represents the 95\% bootstrapped confidence region of the fit. }
\label{fig:z3_LF}
\end{figure}

\begin{figure*}
\centering
\includegraphics[height=5in,width=7in]{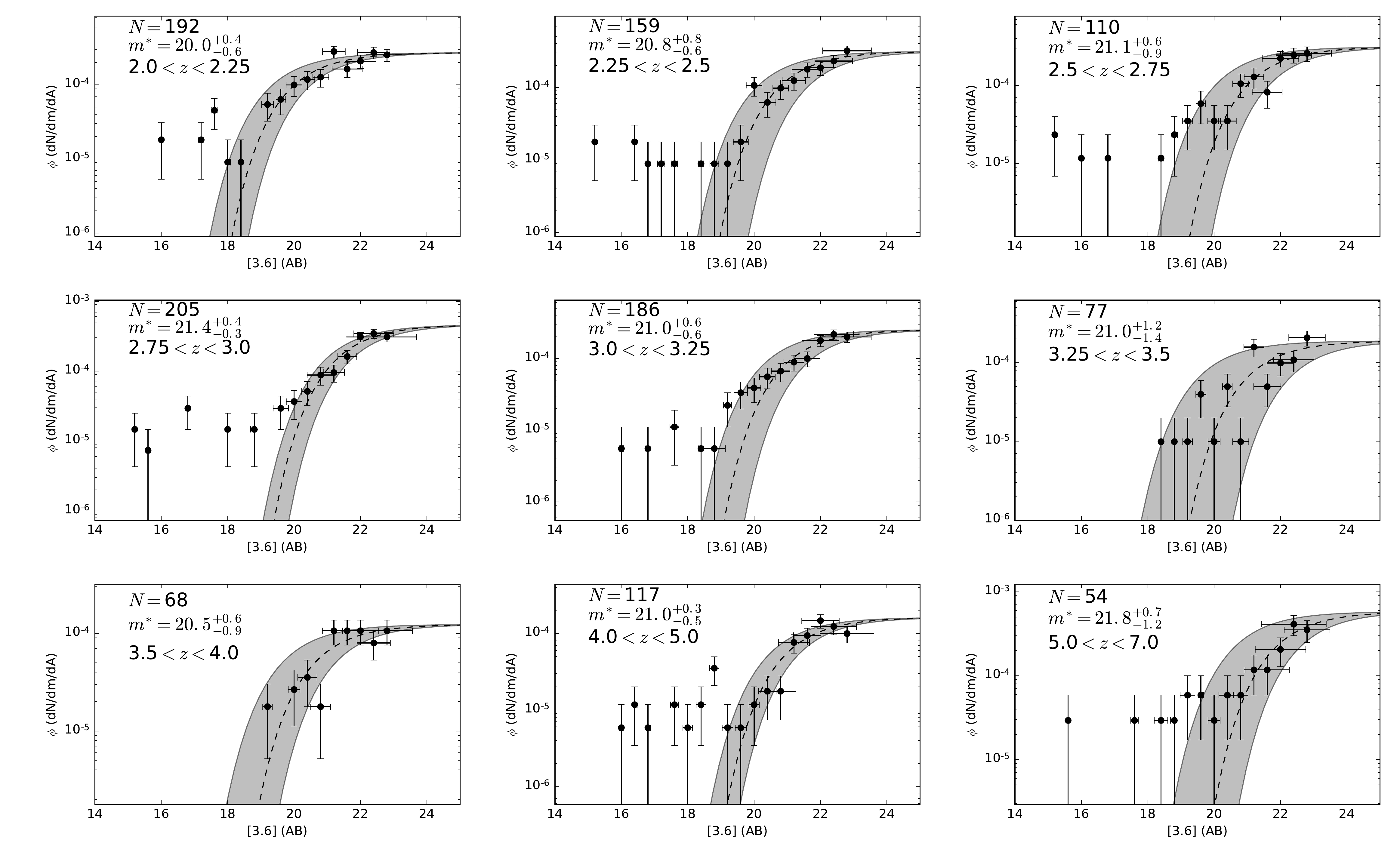}
\hfill
\caption{The 3.6$\mu$m \Spitzer Luminosity Function at nine epochs across $2\le z<7$. The surface number density ($\phi$)
of galaxies is measured in square arcminutes. The number density uncertainty is calculated by the 
root $N$ value in each bin, while the magnitude uncertainties are the average photometric errors in that bin.
 The gray shaded regions are the 95\% confidence intervals calculated by bootstrapping.
The Schechter fit (Eq~\ref{eq:LF}) to the CCPC data is the black dashed line, with $\alpha$ defined as -1. The number of galaxies 
and the value of $m^*$ are listed in the upper left hand corner. The $m^*$ uncertainties are the bootstrapped $2\sigma$ values. Galaxies 
much brighter than $m^*$ we refer to as `hyperluminous' sources and will be discussed in Section ~\ref{sec:dis}. }
\label{fig:CH1_TOT}
\end{figure*}

\begin{figure*}
\centering
\includegraphics[height=5in,width=7in]{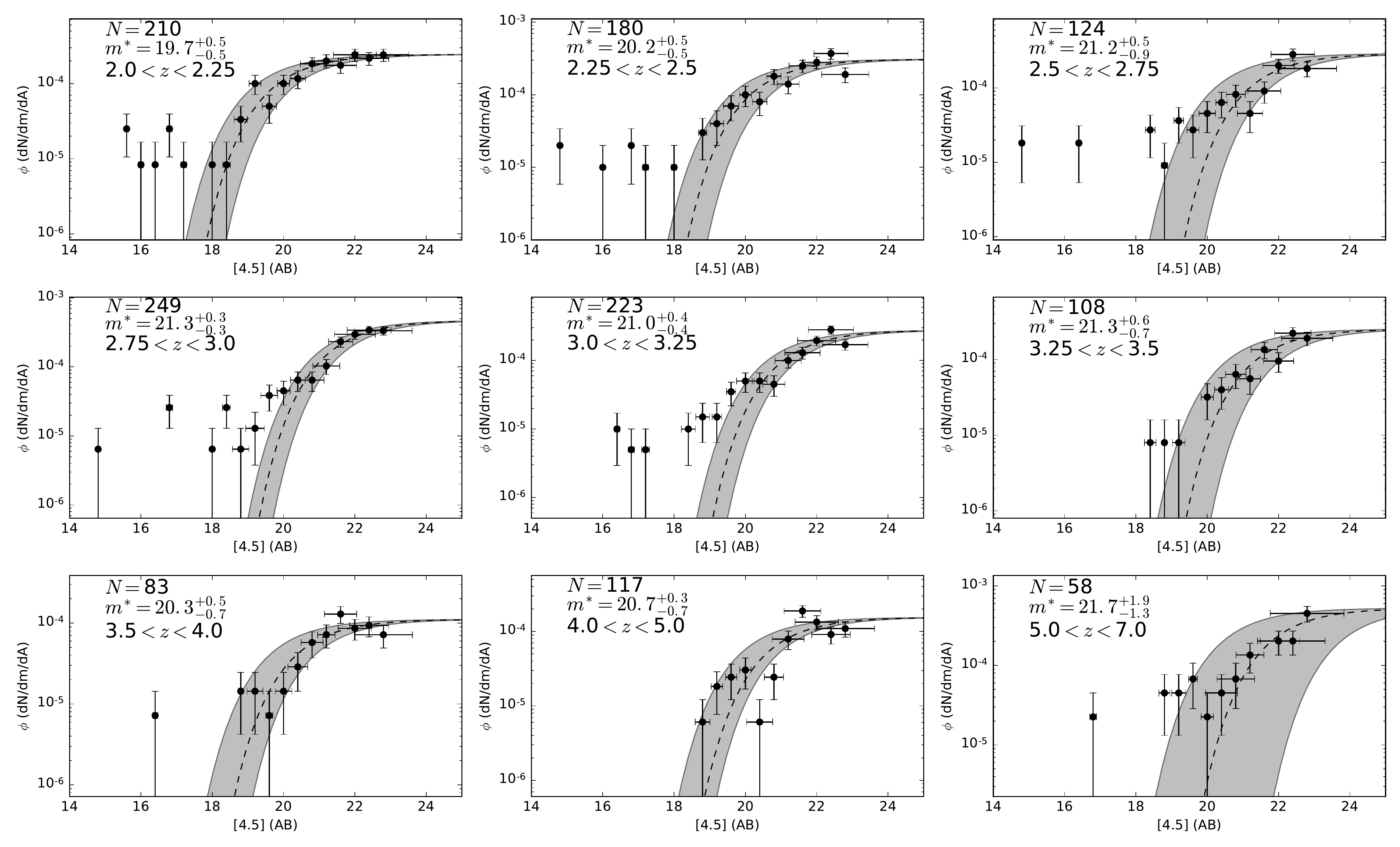}
\hfill
\caption{The 4.5$\mu$m version of Figure ~\ref{fig:CH1_TOT}.}
\label{fig:CH2_TOT}
\end{figure*}

\begin{deluxetable}{c c c c c}
\tablewidth{0pt}
\tablecolumns{5}
\tablecaption{3.6$\mu$\textrm{\normalfont m} CCPC Luminosity Function}
\tablehead{
\colhead{Redshift}	&	\colhead{$N$}	&	\colhead{$m^*$}	&	\colhead{$2\sigma (m^*)$ }	&	\colhead{$\phi^{*}$}		\\
\colhead{Range}	&	\colhead{Galaxies} &		\colhead{(AB)}	&	\colhead{(95$\%$ CI)}	&\colhead{(dN/dm/dA)} 		
}										
$2\le z<2.25$	&	192	&	20.01	&	$_{-0.65}^{+0.50}$	&	2.94$\times10^{-4}$	\\
$2.25\le z<2.5$	&	159	&	20.89	&	$_{-0.68}^{+0.88}$	&	3.41$\times10^{-4}$		\\
$2.5\le z<2.75$	&	110	&	21.13	&	$_{-0.96}^{+0.67}$	&	3.39$\times10^{-4}$		\\
$2.75\le z<3$	&	205	&	21.42	&	$_{-0.33}^{+0.46}$	&	5.01$\times10^{-4}$		\\
$3\le z<3.25$	&	186	&	21.06	&	$_{-0.65}^{+0.60}$	&	2.74$\times10^{-4}$		\\
$3.25\le z<3.5$	&	77	&	21.07	&	$_{-1.47}^{+1.22}$	&	2.05$\times10^{-4}$		\\
$3.5\le z<4$	&	68	&	20.51	&	$_{-0.99}^{+0.63}$	&	1.34$\times10^{-4}$		\\
$4\le z<5$	&	117	&	21.06	&	$_{-0.49}^{+0.38}$	&	1.75$\times10^{-4}$		\\
$5\le z<6.6$	&	54	&	21.83	&	$_{-1.23}^{+0.74}$	&	6.30$\times10^{-4}$		
\enddata
\tablecomments{The results of fitting a Schechter function (Equation ~\ref{eq:LF}) to the CCPC galaxies in a series of redshift bins (1$^{st}$ column).
The number of galaxies ($N$) in each redshift bin is listed in the 2$^{nd}$ column, followed by the fitted $m^*$ parameter 
in the 3$^{rd}$ column. The 4$^{th}$ column represents the 95$\%$ confidence interval of the $m^*$ value. This uncertainty was computed
by bootstrapping with resampling of the data. The characteristic density ($\phi^{*}$) in units of the number of galaxies per magnitude bin 
per square arcminute is found in the final column. The faint end slope of the 
LF was fixed to be $\alpha=-1$.
} \label{tab:ch1_LF}
\end{deluxetable}

\begin{deluxetable}{c c c c c}
\tablewidth{0pt}
\tablecolumns{5}
\tablecaption{4.5$\mu$\textrm{\normalfont m} CCPC Luminosity Function}
\tablehead{
\colhead{Redshift} & \colhead{$N$}	&	\colhead{$m^*$}	& \colhead{$2\sigma (m^*)$ } &	\colhead{$\phi^{*}$}		\\
\colhead{Range} &	\colhead{Galaxies}&	\colhead{(AB)}	&\colhead{(95$\%$ CI)}	&	\colhead{(dN/dm/dA)} 	
} 
\startdata
$2\le z<2.25$	&	210	&	19.73	&	$_{-0.59}^{+0.56}$	&	2.65$\times10^{-4}$		\\
$2.25\le z<2.5$	&	180	&	20.27	&	$_{-0.53}^{+0.56}$	&	3.35$\times10^{-4}$		\\
$2.5\le z<2.75$	&	124	&	21.25	&	$_{-0.97}^{+0.58}$	&	3.14$\times10^{-4}$		\\
$2.75\le z<3$	&	249	&	21.36	&	$_{-0.31}^{+0.39}$	&	5.13$\times10^{-4}$		\\
$3\le z<3.25$	&	223	&	21.08 &	$_{-0.46}^{+0.41}$	&	3.00$\times10^{-4}$		\\
$3.25\le z<3.5$	&	108	&	21.31	&	$_{-0.80}^{+0.71}$	&	2.76$\times10^{-4}$		\\
$3.5\le z<4$	&	83	&	20.38	&	$_{-0.74}^{+0.51}$	&	1.21$\times10^{-4}$		\\
$4\le z<5$	&	117	&	20.71	&	$_{-0.74}^{+0.35}$	&	1.68$\times10^{-4}$		\\
$5\le z<6.6$	&	58	&	21.75	&	$_{-1.38}^{+1.97}$	&	5.69$\times10^{-4}$			
\tablecomments{Identical to Table~\ref{tab:ch1_LF}, but at 4.5$\mu$m.
} \label{tab:ch2_LF}
\end{deluxetable}

Table~\ref{tab:ch1_LF} and Table~\ref{tab:ch2_LF} contain the [3.6] and [4.5] \Spitzer parameters estimated for the CCPC, respectively.
In general, the values of $m^*$ are flat as a function of redshift, suggesting little evolution. 
Figure ~\ref{fig:z3_LF} shows a Luminosity Function for a single epoch to illustrate the finer details of the fit.
Figure ~\ref{fig:CH1_TOT} shows the full set of Luminosity Functions at each epoch at 3.6$\mu$m, while Figure ~\ref{fig:CH2_TOT}
does the same for [4.5].

\subsection{Field Luminosity Functions}

\begin{figure*}
\centering
\includegraphics[height=5in,width=7in]{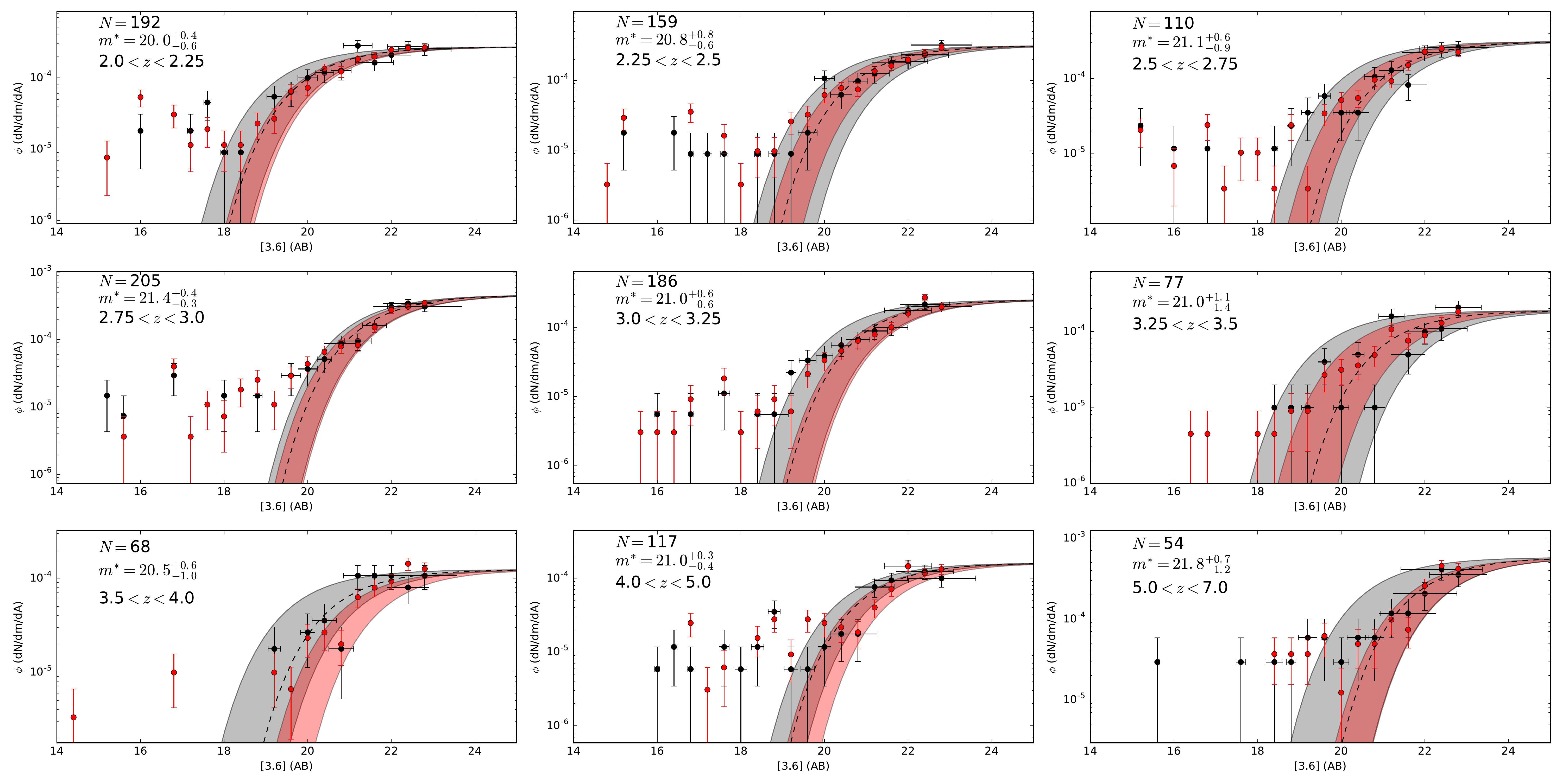}
\hfill
\caption{Superimposed on the LF of CCPC galaxies (Fig~\ref{fig:CH1_TOT}) are spectroscopic `field' galaxies (red points), taken
from the same redshift source catalogs and fields as the CCPC candidates (see Appendix). These are simply scaled
to the CCPC $\phi$ values by their relative numbers at each redshift bin ($\phi_{field}(z)=(N_{CCPC}/N_{field})\phi_{CCPC}(z)$). Within
the uncertainties, these distributions show no statistically significant difference. The red shaded regions are
the bootstrapped 95$\%$ confidence interval for the field galaxies, and they overlap the CCPC's $2\sigma$ range (gray shaded region) at all times. }
\label{fig:CH1_TOT_FIELD}
\end{figure*}

\begin{figure*}
\centering
\includegraphics[height=5in,width=7in]{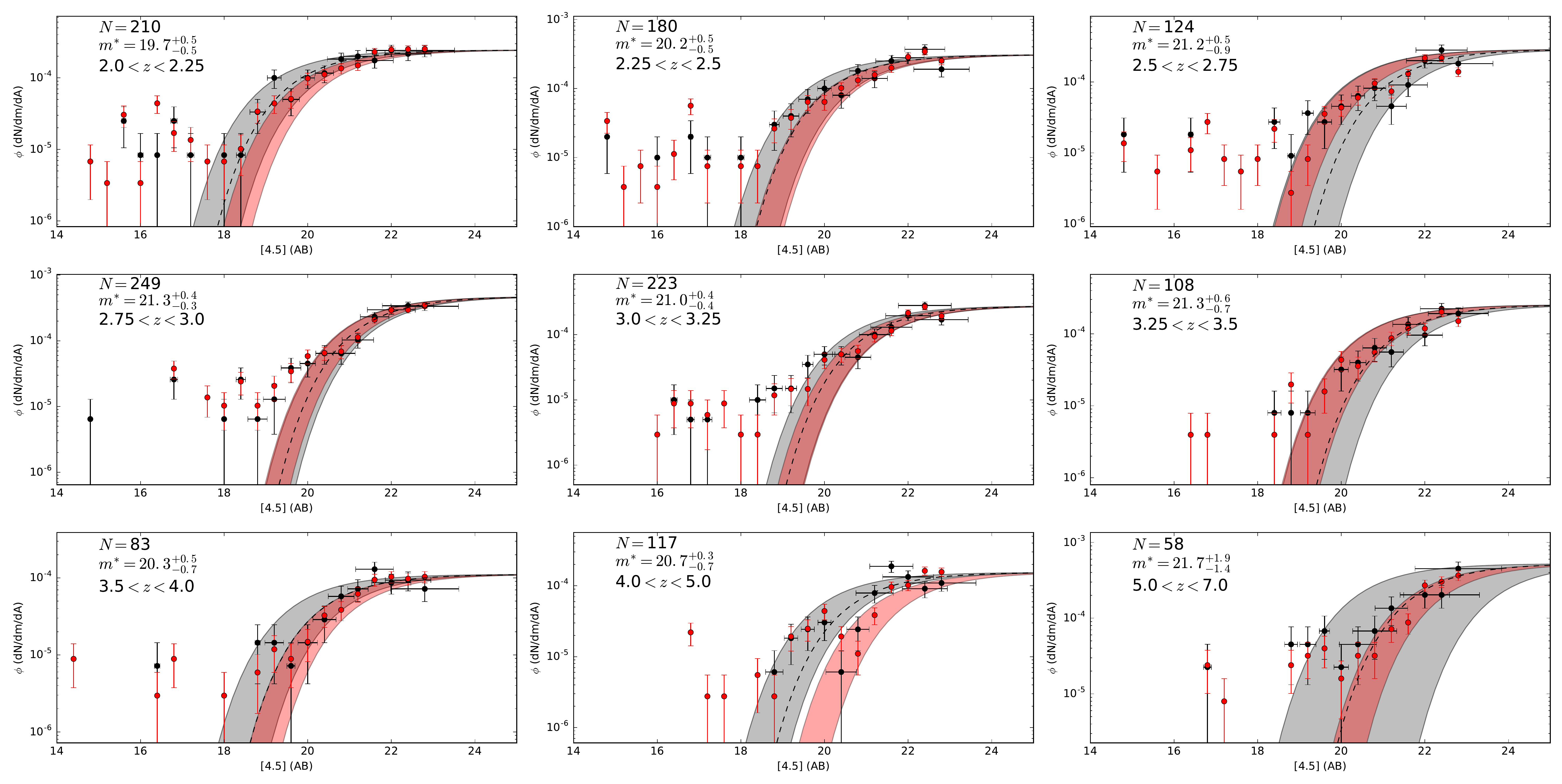}
\hfill
\caption{The same as Fig~\ref{fig:CH1_TOT_FIELD} but with [4.5] magnitudes. }
\label{fig:CH2_TOT_FIELD}
\end{figure*}

In an effort to compare galaxies in overdense environments with their `field' 
counterparts, we assembled a list of all spectroscopic galaxies that were in the same surveys 
as the protoclusters in our sample. This list contains more than 4000 galaxies, double the 
number of the CCPC ($N=2048$ objects). We imposed no richness or density criteria on this sample set, 
apart from that its members were not within the volume of a candidate protocluster. Although 
some of these galaxies may exist in a volume with $N\ge4$ galaxies (a requirement for CCPC candidacy),
they did not have the sufficient galaxy density to be flagged as a protocluster candidate.
The references to the spectroscopic measurements of the field systems are located
in the Appendix, and can also be found in the references of \citet{ccpc1} and \citet{ccpc2}.

We analyzed these galaxies in the same manner and built an `All Galaxy' LF 
(as a proxy for field galaxies) at each redshift bin. The value of $\phi^*$ was scaled to 
the CCPC LF at each epoch. As can be seen in Figs~\ref{fig:CH1_TOT_FIELD} and ~\ref{fig:CH2_TOT_FIELD}, the number density of 
field galaxies (red points) in all bins, at all epochs, are consistent with the CCPC $\phi(m)$ (black points). Fitting a Schechter 
function to `All Galaxies' produces equivalent values of $m^*(z)$ to that of the CCPC galaxies, as can be seen
by the overlapping 95\% confidence intervals (red and gray shaded regions, respectively). 

The CANDELS GOODS-S field is the deepest, most continuous spectroscopic survey from which dozens of structures are identified in 
our sample \citep{ccpc1,ccpc2}. Nearly 25$\%$ of galaxies in the CCPC originate in this field. To 
minimize the effects of varying spectroscopic selection functions from the heterogeneous sample the 
CCPC is constructed from, we constructed GOODS-S LFs for CCPC and non-CCPC galaxies. The 
$m^*$ values remain unchanged at all redshifts in this subsample, for both field and overdense galaxies,
within the uncertainties. Although spectroscopic selection is not definitively ruled out as a variable
for the entirety of the CCPC LFs, it does not appear to be a driving factor in the CANDELS GOODS-S data. 

As a further test, we limited our analysis to galaxies in the `All Galaxy' list that had $N<4$ galaxies
within the CCPC search volume ($R=20$ cMpc, $\Delta z\pm$20 cMpc). Although this `Reduced Field Galaxy' sample limited the number of galaxies 
to 2299, this list is still larger than the number of CCPC galaxies, and thus remains a fair comparison. Interestingly,
this imparted no measurable difference when compared to the shape of the `All Galaxy' or CCPC LFs. It is possible that 
at these very early epochs, the galaxies that are spectroscopically selected have not had sufficient 
time to interact in their modestly dense environments and thus differentiate themselves. Another 
plausible explanation is that many of the galaxies selected in the overdense volumes are not future
cluster galaxies, but rather are field interlopers that will disperse by $z=0$ and thus show little/no
differentiation from our field sample. These results will be discussed in more detail 
in Section~\ref{sec:dis}.

\subsection{Galaxy Colors}

\begin{figure*}
\centering
\includegraphics[scale=0.75]{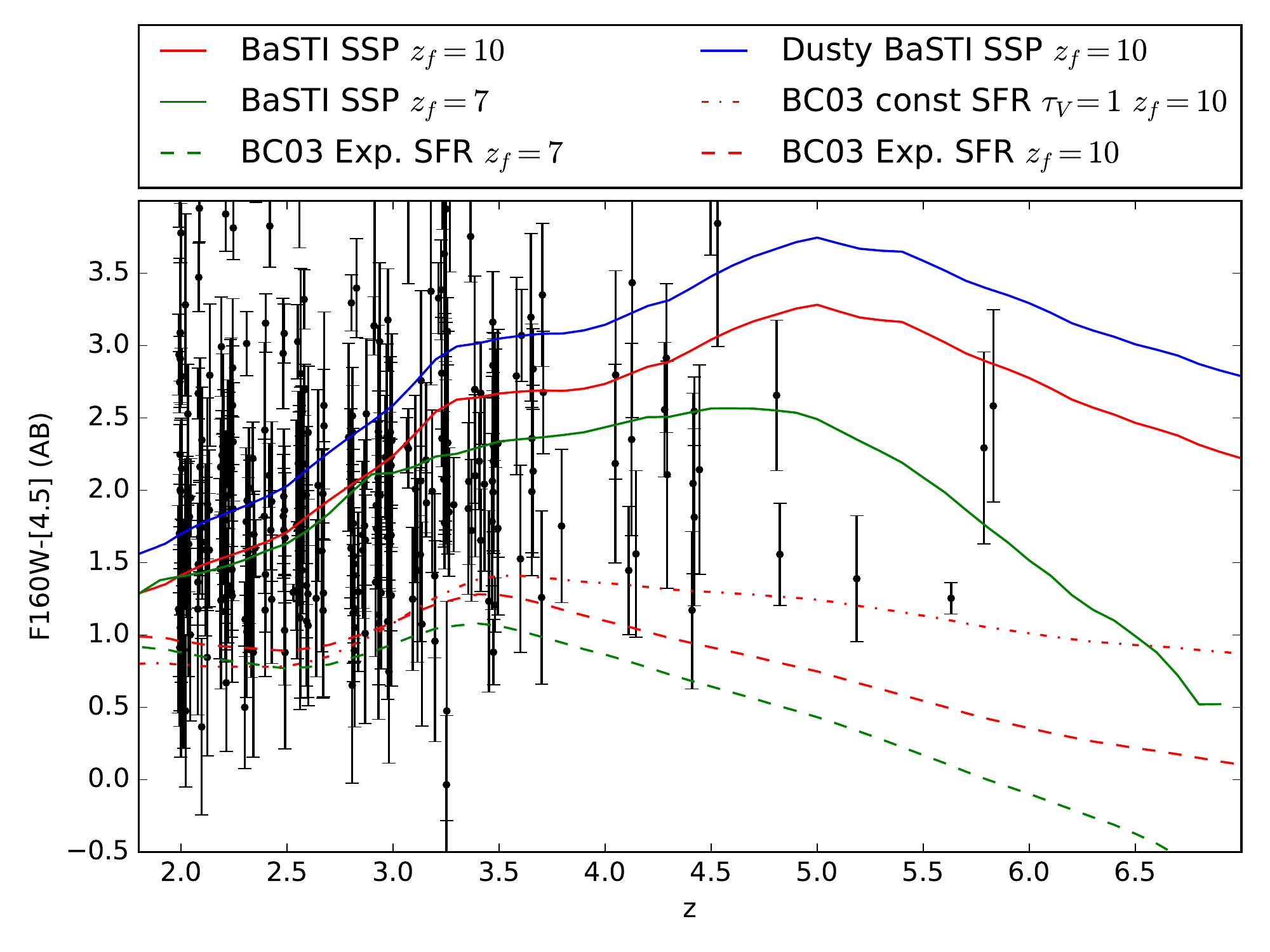}
\hfill
\caption{We plot the $F160W-[4.5]$ color evolution of six stellar populations: two BaSTI SSPs at $z_f=7,10$ (green and red solid lines, respectively), with an added 
\citet{cf00} dust component to the $z_f=10$ SSP (blue solid line), along with two BC03 exponential decaying SFR with $\tau=1$ Gyr at $z_f=7,10$ (green and red dashed lines), and finally
a constant star formation model with an extinction of $\tau_V$ and $z_f=10$ (red dot-dashed line). All models have solar metallicity. Plotted as black
points are the colors of CCPC galaxies. These span the entire range of model stellar populations. For an individual galaxy, 
the uncertainties are generally too large to be assigned to one model or another. However, most systems can be separated between 
SSPs (generally quiescent) and star forming models.}
\label{fig:HST_COL}
\end{figure*}

A selection tool used to identify high-redshift ($z>1.3$) galaxies with \Spitzer is 
the popular $[3.6]-[4.5]> -0.1$ color cut \citep{pap12,wyl14}. It is a simple and 
effective way of removing low redshift galaxies from a sample, regardless of the 
underlying stellar population. Interestingly, more than 1/3 of the CCPC sample 
(all with spectroscopic redshifts $z>2$) failed this color cut. Within the photometric 
uncertainties, many of these systems emit `true' colors that would satisfy the criterion. 
However, this blind cut can remove a significant portion of a sample of high redshift galaxies. This 
is not unique to our photometry. The galaxies in CCPC structures that are coincident with 
objects in the 3D-HST database \citep{ske14} show a similar result, with 33.8\% of more 
than 300 galaxies having $[3.6]-[4.5]< -0.1$, with a median color of $[3.6]-[4.5]= -0.25$. This is 
a further piece of evidence suggesting that this aberrant fraction is the result of 
photometric uncertainty that has scattered the colors below the cut. There is no 
obvious correlation with apparent magnitude, therefore AGN or hyperluminous source contamination is likely not an issue. 

Although effective at measuring the underlying stellar mass of high redshift galaxies
from their rest-frame NIR emission, the colors of \Spitzer are not sensitive to different stellar populations \citep{coo14}.
Even at the highest redshift of the CCPC ($z=6.56$), the rest-frame wavelength observed at 4.5$\mu$m
falls at just 5900\AA. As the majority of these galaxies were spectroscopically targeted as UV bright, star forming systems,
it stands to reason that they should have blue colors, which \Spitzer is not sensitive to. The \HST WFC3 $F160W$ filter 
probes rest-frame wavelengths of $2100-5300\AA$  between the redshift range $2.0<z<6.56$. Figure ~\ref{fig:HST_COL} illustrates 
the redshift evolution of the $F160W-[4.5]$ galaxy color as a function of redshift for a variety of stellar population 
models \citep{bc03,basti}, with different initial mass functions \citep{chab,krou} and dust extinction \citep{cf00}. Plotted
on the models are the $F160W-[4.5]$ colors as a function of redshift for the CCPC. We only show galaxies with 
photometric uncertainties in an individual filter of $\sigma<0.75$ mags. There is no clear preference
for a single galaxy type, with a large scatter of blue and red galaxies throughout. We also cannot ascertain a preferred
stellar population, dust content or formation redshift from these colors alone for an individual galaxy, as the uncertainties 
can be too large. For many systems the difference between a star forming galaxy and a passively evolving SSP can be 
assigned within the errors.

\citet{noi16} spectroscopically targeted two candidate structures identified by \citet{wyl14} at $z\sim2$. Their \HST
photometry of the overdensity revealed a wide spread of F140W-[3.6] colors of roughly $-1 < F140W-[3.6] < 3$, with a much 
smaller range for the galaxies with confirmed redshifts (approximately $0.5 < F140W-[3.6] < 2$). 

The 3D-HST data \citep{ske14}
coincident with CCPC galaxies in the CANDELS fields at $2.0<z<2.05$ has a mean color of $<F140W-[3.6]>=0.5\pm0.6$, and a range
$-0.4 < F140W-[3.6] < 2.3$. The color range is more broad than the spectroscopic sample of \citet{noi16},
but does not go beyond the full range of candidate galaxies in their protoclusters. This suggests a general agreement
of their protocluster galaxy population and ours.
Higher redshift CCPC galaxies show little color evolution in this plane as well.



\subsection{SED Fitting and Galaxy Stellar Mass}

\begin{figure*}
\centering
\includegraphics[height=4.0in,width=7.0in]{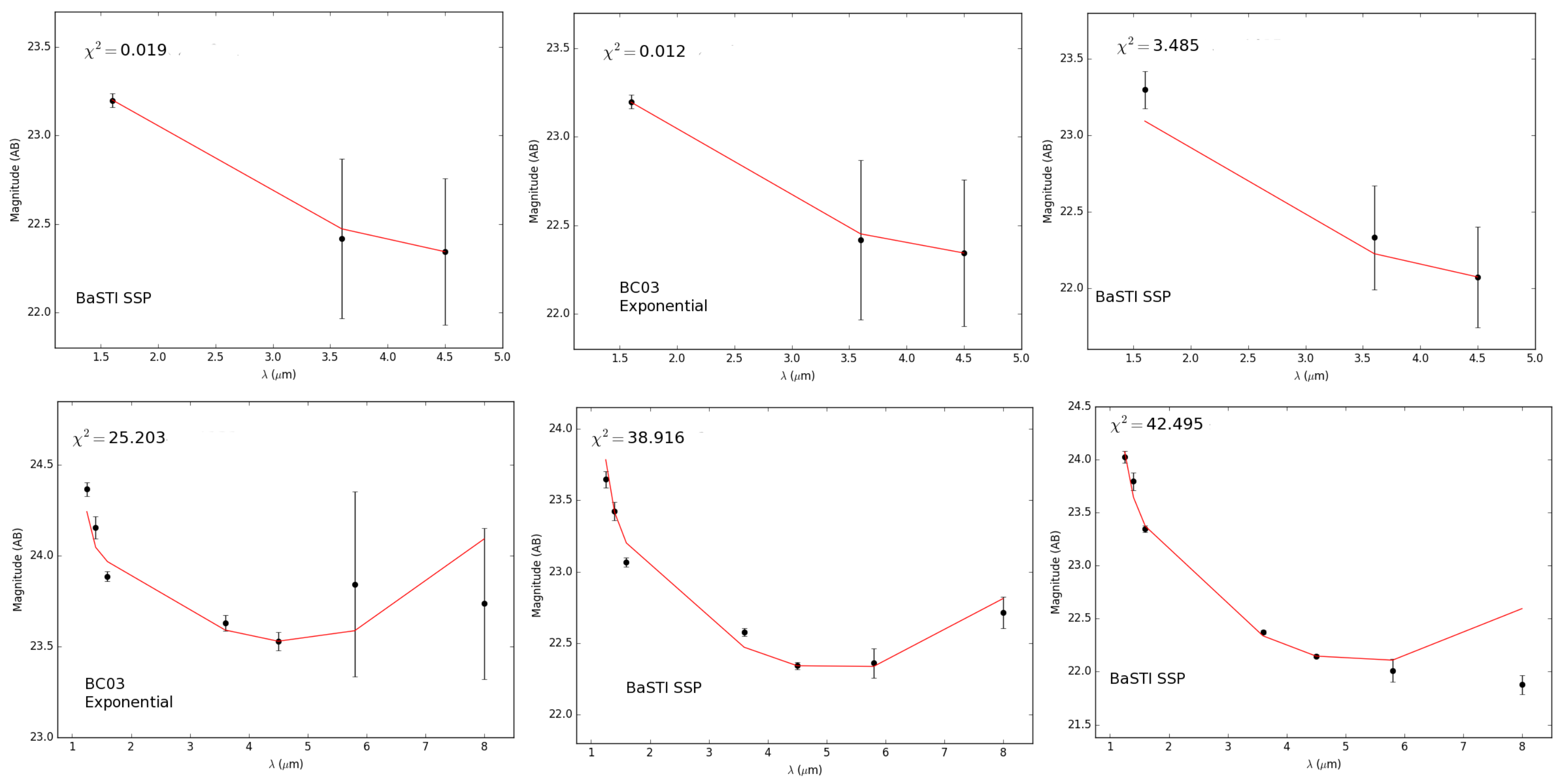}
\hfill
\caption{\emph{Top:} Three examples of CCPC galaxy SED fits. The black points are the apparent magnitudes 
at observed wavelengths (F160W, [3.6], and [4.5] filters), while the red lines are the template magnitudes. The formation redshifts, from left to right,
are $z_f=3.0$, 7.0, and 3.25. The left and right templates are BaSTI SSPs with solar metallicities, while the center template is a BC03, 
exponentially decaying model with $\tau=1$ Gyr. The $\chi^2$ values of the fits are listed, which are used to distinguish 
the relative goodness-of-fit between various models for a single galaxy. \emph{Bottom:} CCPC galaxies found within the 3D-HST database \citep{ske14} can have 
greater wavelength coverage (F125W, F140W, F160W, [3.6], [4.5], [5.6], and [8.0]), although these are limited to the GOODS fields. 
The templates presented are dust-free. The left panel is a BC03 exponential model with $\tau=1$ Gyr and metallicity of
$Z=0.008$, while the middle and right panels are BasTI SSP with solar metallicities. They have $z_f=3.75,$ $3.0,$ and 3.5 from left to right, respectively.
\emph{The larger $\chi^2$ values of the models in this panel compared to the \emph{Top} panel are the result of more
measurements (with relatively small uncertainties) from which a galaxy model can differ. They are not necessarily a poorer fit. 
The $\chi^2$ values also depend upon the inherent biases within the models as well, and should therefore not 
be treated as an absolute metric.}  \label{fig:SED}}
\end{figure*}

The usual methodology of obtaining stellar masses for galaxies at high redshift
is to match a stellar population model with a number of observed magnitudes across a 
range of filters \cite[e.g. spectral energy distribution (SED) fitting;][]{mag10}. Thus, a galaxy is matched with the stellar age,
star formation history, metallicity, initial mass function, and dust content of a given 
model. Hereafter we will refer to this as `traditional' SED fitting.
 The implied stellar mass-to-light ratio of the best fit model is then used to compute the $M_{\star}$ of the galaxy.
There are a number of degeneracies that arise from this method. This can be particularly important with
limited wavelength coverage and inconsistent rest-frame colors over a
large range in redshift. These factors will be briefly considered in Section ~\ref{sec:dis}.

To test the relative strength of a fit to different model SEDs, we follow
the procedure used in \citet{hyperz} to estimate photometric redshifts:
\begin{equation}\label{eq:chi}
\chi^2 = \sum\limits_{i=1}^{N_{filt} } \frac{ (F_{i,obs}-F_{i,model})^2 } { \sigma_i^2 } 
\end{equation}
which measures the observed flux in a filter ($F_{i,obs}$) relative to the flux of a model stellar
population ($F_{i,model}$) and weighted by the squared observational flux uncertainty ($\sigma_i^2$).
The SED's flux is calibrated to match the observed [4.5] magnitude of the galaxy being measured in our algorithm.
We limit our analysis to magnitudes with uncertainties $\sigma_m < 0.75$ and require a F160W measurement. As stated previously, \Spitzer
colors cannot differentiate between varying stellar populations at these redshifts. This maximum allowable uncertainty is effective 
in limiting severely anomalous photometric measurements from a heterogeneous sample of surveys which do not have a constant depth. We will show that 
the photometric uncertainties are generally of minor importance when compared to the 
model degeneracies of our fits. The stellar mass implied by different models 
for the same redshift and magnitude have greater variance than the uncertainty
introduced by photometric errors. This will be discussed further in Section ~\ref{sec:dis}.


Using the EzGal code \citep{ezgal}, we built grids for each filter and stellar population
model as a function of formation redshift ($2<z_f<10$) and observed redshift ($z_{obs}$). We consider
 the following models: BaSTI simple stellar populations (SSPs) with metallicities of $Z=0.008,0.0198$ \citep{basti}
and a Kroupa IMF \citep{krou}, \citet{bc03} (hereafter BC03) constant star formation models with extinctions of 
$\tau_V=0.2,1.0$ and a \citet{chab} IMF, and two BC03 exponential decaying SF models with $\tau=1.0$ Gyrs and 
$Z=0.008,0.02$ . The model predictions become erratic when the stellar age of the system is low, so we implemented
a cut of $z_f -z_{obs} > 0.05$ when fitting the formation redshift. 

For the models that did not have a dust component built in, we also explicitly calculated the 
extinction from a \citet{cf00} model for the rest-frame wavelengths observed in our filters at a stellar age
computed from the $z_f$ value in the grid. The dusty fluxes were recomputed and the fits measured using Equation 
~\ref{eq:chi} in the same manner as the dust-free systems. 

Once the best-fitting model is identified for each galaxy, we calibrate the model to the observed magnitude $m_i$, for each filter
measured ($i$), and query EzGal for the implied stellar mass at the observed redshift. For an individual filter's mass measurement, $M_{i,\star}$, we estimate its uncertainty by taking the
implied mass of the galaxy if the magnitude was changed by its photometric uncertainty ($m_i\pm \sigma_{i} $) in that filter. Ultimately, the
estimated stellar mass of the system is computed from the uncertainty-weighted mean value from each wavelength measured.
If the system is found to have a best-fit SED that is dusty, we subtract the dust absorption from the observed magnitudes
prior to computing the underlying stellar mass estimates. 

In this manner, we were able to fit 414 galaxies that were below a minimum $\chi^2$ threshold, with a median value of $\chi^2 = 1.6$ 
and a median of $N_{filters}=3$ (F160W, [3.6], and [4.5]). Fig~\ref{fig:SED} shows a few examples of SED fits with 
a variety of $\chi^2$ values. The median mass implied for these galaxies is 3.3$\times10^{10}$ \Msun. Nearly half 
of the systems were best fit by an exponentially decaying, BC03 star forming model (185 objects),
with 201 others well fit by a BaSTI simple stellar population. 
Roughly $50\%$ of galaxies were found to be best fit by a dusty component, and only 
18\% had less than solar metallicity. The average formation redshift fit by the algorithm was generally
old, at $z_f=7.9$. The 3D-HST data \citep{ske14} has greater wavelength coverage in \HST filters within the CANDELS fields. 
The fits to 395 SEDs using this expanded data set did have a marginally smaller median $M_{\star}$ value of 0.9$\times10^{10}$ \Msun, 
a lower percentage of dusty galaxies (27\%), and 51\% low-metallicity systems (compared to $<15\%$ in our data set). The
3D-HST catalog's photometry did not use aperture corrections in their \Spitzer magnitudes, which (if instituted) would systematically
increase the stellar mass of these galaxies.  Only 7\% of the 3D-HST galaxies were best fit by a BaSTI SSP, while the BC03 constant star 
formation models was applied to roughly 1/3 of CCPC sources. The majority were fit by an exponential model, just
like our own photometric set. The mean formation redshift for the 3D-HST photometry was $z_f=5.9$.
The 3D-HST fits have a median value of $\chi^2 =$30.5.

With the recent success of modeling galaxy SFHs as a log-normal distribution \citep{gla13,abr16}, we 
attempted to fit the CCPC with a similar analysis. The log-normal distribution adopted is of the form
\begin{equation}\label{eq:logn}
SFR\propto \frac{\exp{(-\frac{\big(\ln(t)-T_0)^2}{2\tau^2}\big)	}}{t}
\end{equation}
where $t$ is the time since the Big Bang, $T_0$ is the half-mass time of the galaxy, and $\tau$ is the half-mass
width of the distribution \citep{abr16}. We took the two BaSTI SSPs (different metallicities) as our base models, 
and then computed Complex Stellar Populations (CSPs) using EzGal for a variety of values of $T_0,\tau$ ranging from
 $0.05\le(T_0,\tau)\le1.0$. This range corresponds to the breadth of values fit to observed SEDs at low 
and high redshifts \citep{gla13}. We evaluated the goodness-of-fit with Equation ~\ref{eq:chi}, as before. We included the
optional \citet{cf00} dust extinction by assuming a stellar population age ($z_f$) coincident with a SFR$(t)=10^{-3}($SFR$_{max}$). 
The factor $10^{-3}$ is fairly arbitrary, with an order of magnitude adjustment changing $z_f$ by approximately 0.2. 

The log-normal routine was able to adequately describe only 207 galaxies, roughly half of the traditional number of successful
SED fits. The fits were also poorer than the traditional fitting, with a median $\chi^2 = 24.6$ (compared to $\chi^2 = 1.6$).
This is somewhat surprising, as the volume of parameter space explored was much larger than in traditional fitting.
The median mass for these galaxies (1.5$\times10^{10}$ \Msun) was similar to the traditional fit, with no  
preference for low metallicity systems (15\%). Nearly 85\% of the galaxies were best fit with a dust component included.
The mean formation redshift was $z_f=10.7$, only slightly older than the stellar age from traditional fitting. The median 
values of $T_0,\tau$ are 0.05 and 0.15 ($\sim$1.05 and 1.16 Gyrs), respectively. Running the SED fitter on the 3D-HST data set generally confirms the earlier
results from our photometry, with a few notable exceptions. Interestingly, more galaxies were successfully fit by using 
the larger wavelength coverage (262) but with a much lower fraction of dusty systems (0.33), a larger mean $T_0$ (0.65), an older
mean stellar population ($z_f=14.6$), and a slightly lower
median mass than the traditional models (0.7$\times10^{10}$ $\Msun$). The median values of $\tau=0.15$ was equivalent 
to the log-normal value derived with our photometry. The 3D-HST fits had a similar value of $\chi^2=21.7$ to the log-normal fit of our data,
but are in fact more robust, as more filters were used. We will also briefly note
that although the median values here may appear to be slightly different if traditional or log-normal SED fitting is used, or
the 3D-HST data adopted versus our own photometry, the variance of the properties are much larger than their differences. This suggests
that the underlying properties of these galaxies are still very uncertain. This will be discussed further in Section~\ref{sec:dis}.

\begin{figure*}
\centering
\includegraphics[scale=0.75]{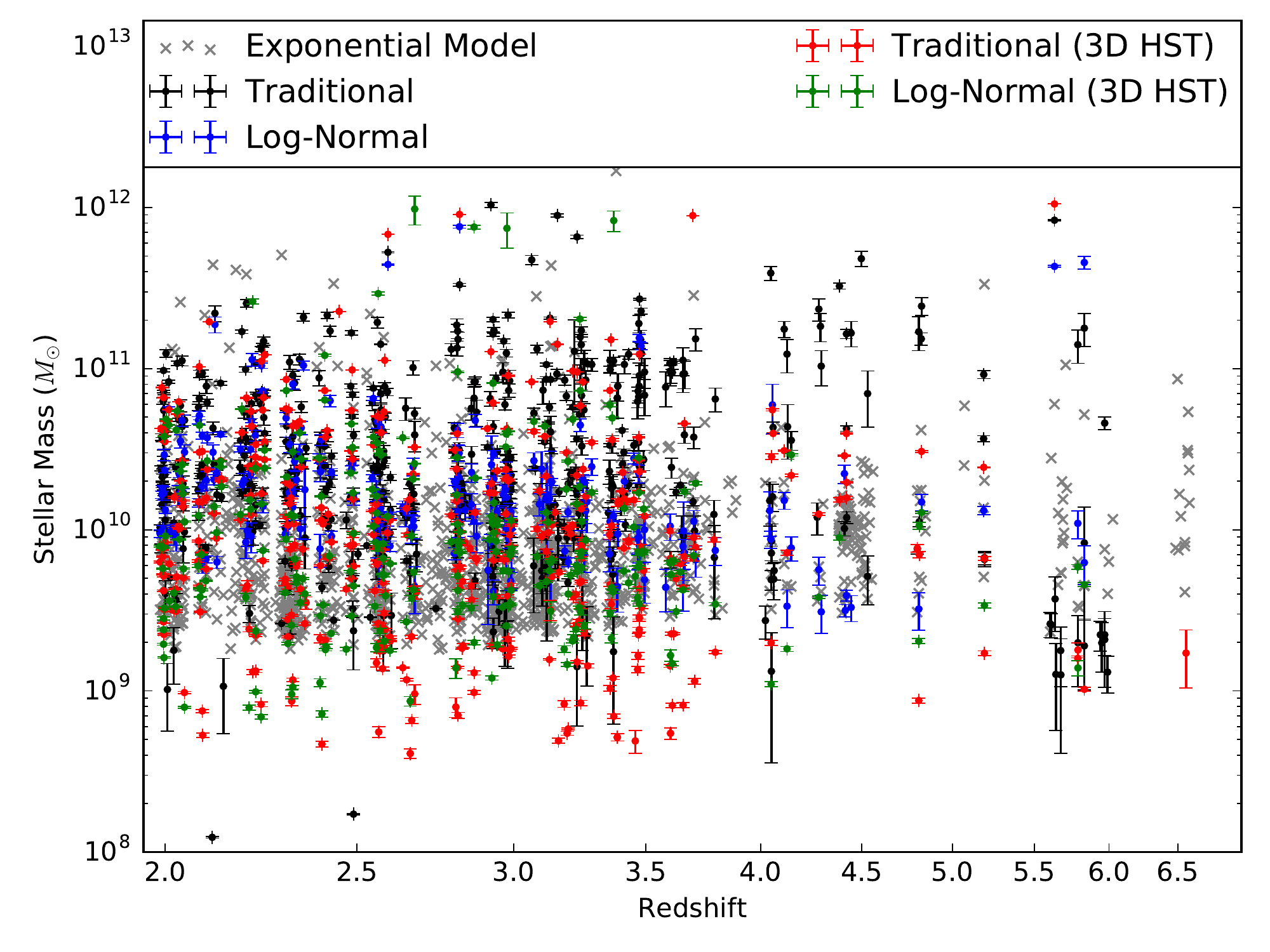}
\hfill
\caption{The stellar masses (in units of \Msun) as a function of redshift (note the logarithmic scale)  for CCPC galaxies estimated via traditional SED
fitting (black points), and log-normal SFH fitting (blue points). All galaxies with a [4.5] measurement are shown as 
gray points, which assume a young, bright stellar 
population with an exponentially decaying SFR. Also included are galaxies photometrically measured by the 3D-HST project
\citep{ske14} by traditional and log-normal SED fitting (red and green points, respectively). Individual galaxies with more 
than one stellar mass estimate can have large variances (much larger than the photometric errors) 
based on the model selected with the minimum $\chi^2$ value.  For instance, a bright galaxy with a young stellar population 
can have an order of magnitude lower stellar mass than the same galaxy fit by an old SSP. We are not able to adequately select a unique model for a given 
galaxy with our data. The masses are the uncertainty-weighted mean values from the adopted population model. The 
gray error bars are suppressed for clarity, but are generally larger than the other data points, as they are only measured at 4.5$\mu$m. The galaxies 
with the largest stellar masses are typically the `hyperluminous' sources seen in in Figs ~\ref{fig:CH1_TOT} and ~\ref{fig:CH2_TOT}, as expected. 
These include a probable mix of AGN, low-redshift interlopers,
and simply massive systems. }
\label{fig:mass}
\end{figure*}

Many galaxies have no \HST photometry in the CCPC. In order to estimate their $M_{\star}$
we adopt the simple BC03 exponential decaying SF model with a metallicity of $Z=0.02$, $\tau=1$ Gyr, and $z_f=z+0.25$. This choice is reasonably justified, as the 
majority of the CCPC galaxy sample was selected as unobscured star forming systems (UV-bright LBGs or $Ly\alpha/H\alpha$ line-emitters),
and this exponential model appears to be a consistent fit with the evolution of $m^*$. It supplies a conservatively low
mass estimate for the system. 

Figure~\ref{fig:mass} plots the galaxy stellar masses of the objects measured by various SED fitting methods. The scatter is large, even 
for a single galaxy fit by different algorithms (log-normal versus traditional) or different photometry (3D-HST versus our own).
The gray points are the conservative mass estimate for each galaxy, which is simply an exponential model (BC03, $Z=0.02$, and $z_f=z+0.25$ 
described above). These points can be thought of as lower limits of stellar mass for individual galaxies. An example presented in 
Section~\ref{sec:dis} will illustrate the circumstances in which this is the case.  

\section{Discussion} \label{sec:dis}

\subsection{Hyperluminous Sources}

It is clear that the Schechter function fits the data adequately within a few magnitudes around $m^*$. However, there are
some very bright sources ($m \le m^* -2$) that are clearly anomalous in Figures ~\ref{fig:CH1_TOT} and ~\ref{fig:CH2_TOT}. These points are rare, generally consisting of 1-3 
galaxies per bin, but they are apparent at all redshifts and at both wavelengths. In the context of \LCDM and hierarchical accretion,
it is predicted that in the densest regions of the universe at high redshift ($z>2$), the most massive galaxies will reside \citep{mul15}.
Therefore, protocluster galaxies might be expected to be in the most massive halo systems. Curiously, field galaxies appear to have the same
proportion of hyperluminous galaxies as their overdense counterparts (Figs ~\ref{fig:CH1_TOT_FIELD},~\ref{fig:CH2_TOT_FIELD}). We will discuss a few other possibilities 
for the origin of these sources.

In their Fig 23, \citet{guo11} plot their semi-analytic model (SAM) Schechter stellar mass functions at redshifts $2<z<4$, overplotted with
observational data from \citet{per08} and \citet{mar09}. The data diverge from those models in much the same manner
as the data presented here, with a number of bright objects not fit by the exponentially declining number density.
At a redshift of $z\approx7$, \citet{bow14} find a similar trend at rest-frame UV wavelengths.
It could be possible that at these epochs, Eq~\ref{eq:LF} is not representative of the stellar mass of galaxies.
There is some indication that the most massive galaxies observed at very high redshifts ($z>4$) had not the time to assemble 
in a \LCDM universe \citep{ste16}, and their halo mass density is larger than 
theoretical predictions. Some of the hyperluminous sources may be these galaxies and their descendants.

AGN can have strong, non-stellar emission that dominates the flux of the galaxy. These objects are contributing to the number of hyperluminous sources. 
Roughly half of the hyperluminous objects were spectroscopically selected as part of quasar and AGN surveys, and additionally some 
of our data come from targeted overdensities surrounding these types of sources, as in \citet{ven07}. There is also some 
evidence that AGN are found in greater density surrounding protoclusters \citep{cas15}.
However, AGN are quite rare in LBG studies \citep{mag10}.

Some of these objects were detected using NB filters centered on redshifted $H\alpha$ or $Ly\alpha$ lines, 
which were then confirmed to be emission lines spectroscopically. However, emission line galaxies can be incredibly 
faint and show little or no continuum emission \citep{fyn03}, and thus no other absorption or other emission features are identified.
Therefore, some of these objects could be [O II] emitters ($\lambda\lambda3726, 3729$), or other line-emitting galaxies at lower redshifts,
and are therefore less distant than expected. \citet{ven07} discuss various tests than can be used to disqualify candidate 
$Ly\alpha$ systems, and they estimate that interlopers are $\le10\%$ in their sample. These low redshift interlopers could account for a 
few of these hyperluminous sources.

\subsection{Field Galaxy Comparisons}

More massive galaxy halos are systematically found in denser environments. This is an observed effect at high redshift, where the two-point correlation function 
amplitude appears tied to the UV luminosity of LBGs \citep{ouc04}. It also has a theoretical basis found within large \LCDM simulations, where the most
massive galaxies reside almost exclusively in the densest environments \citep{mul15}. However, it is readily apparent in Figs ~\ref{fig:CH1_TOT_FIELD},~\ref{fig:CH2_TOT_FIELD} that the 
luminosity functions of field galaxies are in no way distinct from their CCPC galaxy counterparts. They contain an equal measure of hyperluminous 
sources, and their respective Schechter function parameters are equivalent. 

Taken at face value, the rest-frame NIR emission of galaxies at all redshifts and densities 
are essentially equal, and therefore their stellar mass contents should be similar as well. In order to reconcile 
this fact with the points laid out in the previous paragraph, we will suggest a few possible solutions.
We note that our data are not sufficient to endorse any of these over another. The null hypothesis is 
that the galaxy stellar populations at $z>2$ are the same, regardless of environment.   

If galaxies in protoclusters are inherently brighter, as expected, but also had a greater 
fraction of dusty galaxies or more dust extinction in general, this could balance the 
magnitude of field and protocluster galaxies. Assuming a \citet{cf00} dust model, 
we can predict the dust extinction for a given wavelength and stellar age. At the redshifts measured,
it is a low extinction of median $\Delta m\sim0.6$ mags for a starbursting galaxy and $\Delta m\sim0.3$ mag
for an older population from $2<z<4$ at $4.5\mu$m, with the extinction increasing at higher redshifts.
This hypothesis would require an extremely convenient steady increase in dust absorption across the range of rest-frame emission
to account for the stellar mass difference between protocluster galaxies and the field. The maximum extinction during a starburst is hardly
significant ($\Delta m \approx0.6$ mags), and might not be detected within the uncertainties of $m^*$.

Another option is that our selection of `field' sources actually targets 
marginally overdense systems, and are therefore not isolated enough to be different
from the CCPC systems. As a reminder, the initial `Field' sample was composed of galaxies
within the same survey fields as the CCPC galaxies to minimize bias, but were not found
within the same volume as a CCPC candidate. After this sample showed no differentiation, a smaller
subsample was crafted which contained $N<4$ galaxies in the same volume as the CCPC. This also did 
not show any difference in the LFs when compared to the CCPC. Analysis of galaxies limited to the 
GOODS-S survey also showed no statistically distinct difference, suggesting that in this particular 
instance, the myriad of spectroscopic selection functions of the CCPC galaxies did not dilute
a potential signal.

\citet{con15} analyzed zoomed-in protocluster galaxies in a SAM, and found that for galaxies in the region of a protocluster,
but not bound to it at $z=0$, the galaxy properties (color, mass, etc.) were indistinguishable. It is possible these
field galaxies may be similarly camouflaged. A related plausibility is that a large number of interlopers within the 
overdensity volumes mask a detectable differentiation. If, however, the rarest, most massive galaxies form only in the densest regions
of the universe \citep{mul15}, then presumably some evidence of these could be solely evident in the CCPC LFs. 
Recently, \citet{hat16} found evidence suggesting that dense sub-groups in a protocluster at $z\sim1.6$ exhibited differentiation 
with respect to the field, while 2/3 of the member galaxies outside of groups showed no variation.

A further possible, but poor, explanation is that 
the spectroscopic selection of the surveys used are more incomplete in these field regions than in the CCPC volumes, and 
therefore may well be overdense themselves. However, spectroscopic completeness is strongly correlated to flux 
for practical purposes, and we see no difference with galaxy densities brighter than $m^*$. 
At the present time, we do not have a satisfactory explanation for this discrepancy.

\subsection{$m^*$ Evolution}

\begin{figure*}
\centering
\begin{subfigure}
\centering
\includegraphics[height=3.25in,width=3.25in]{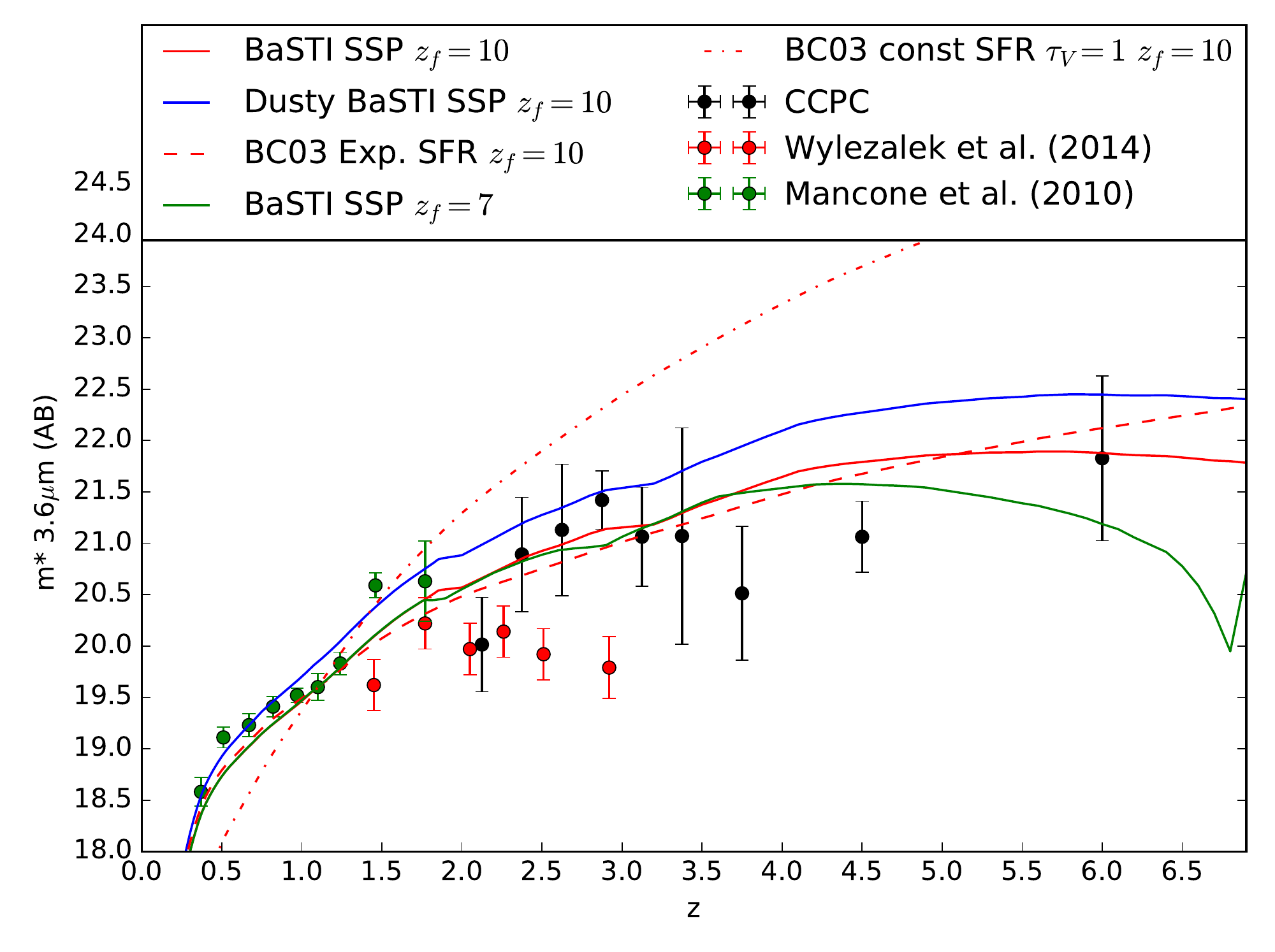}
\end{subfigure}
\hfill
\begin{subfigure}
\centering
\includegraphics[height=3.25in,width=3.25in]{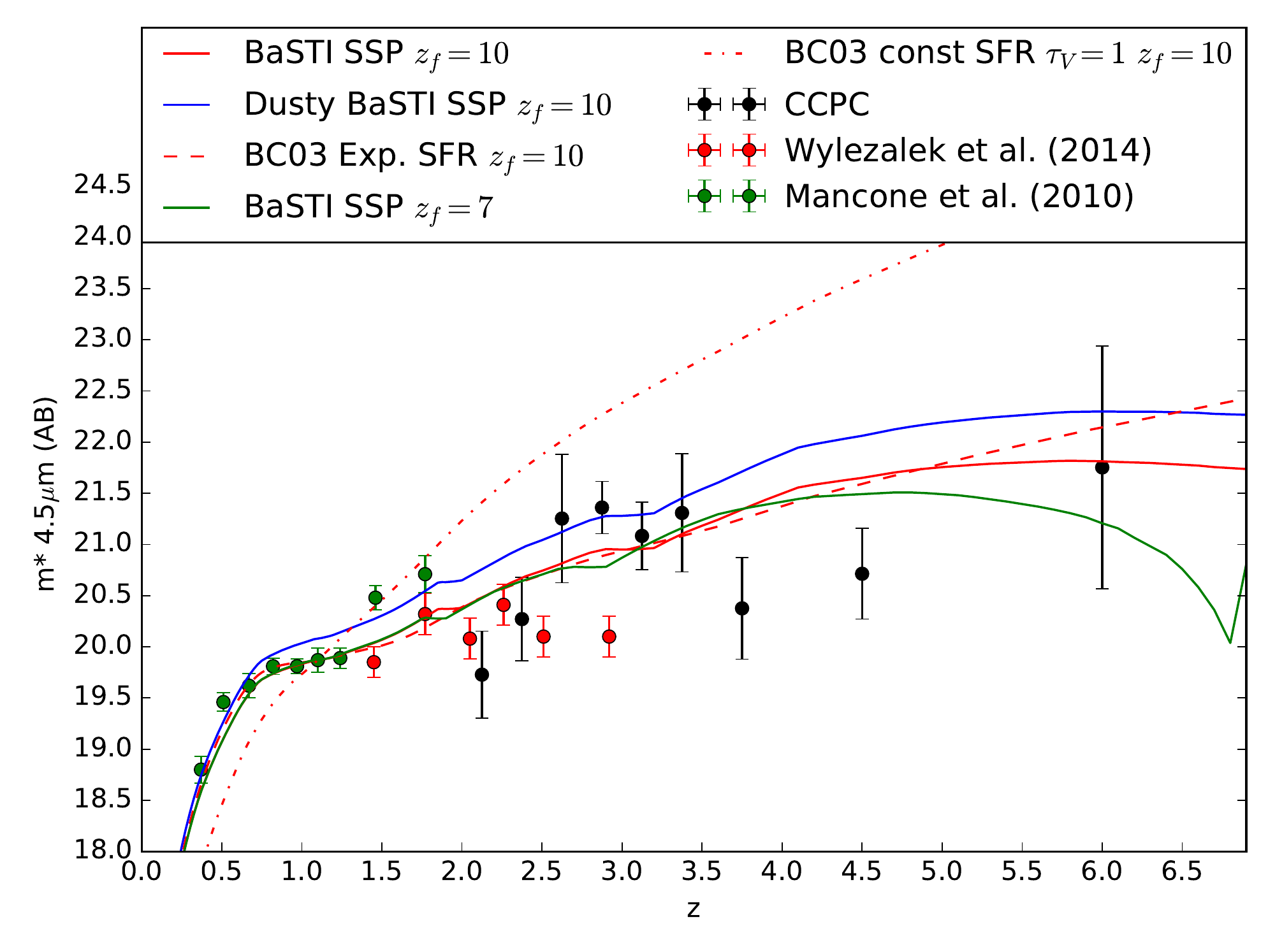}
\end{subfigure}
\hfill
\caption{We show the evolution of $m^*$ as a function of redshift for 3.6$\mu$m (\emph{Left}) 
and 4.5$\mu$m (\emph{Right Panel}). The black points are the CCPC values,
while green and red values are from \citet{man10} and \citet{wyl14}, respectively. The error bars shown for the 
CCPC are $1\sigma$ values computed by bootstrapping.
Overplotted are various models computed using EzGal. Briefly, we have included two BaSTI simple stellar populations at formation redshifts $z_f=7,10$ (solid green and red lines, respectively), and added a \citet{cf00} dust prescription to the $z_f=10$ SSP (blue solid line). From the BC03 models, we included an exponentially declining SFR model with $\tau=1$ Gyr (dashed red line) and a 
constant star forming model with an extinction of $\tau_V=1$ (dash-dot red line). This latter model provides a poor fit to the $m^*$ evolution, while the two SSPs ($z_f=7,10$) and exponential model
are consistent with the data. Each model is scaled to match the $z=1.1$ $m^*$ value from \citet{man10}, although this choice is not unique. The scaling simply
moves the evolution curves to brighter or fainter magnitudes, but the shape remains constant. Therefore, for any scaling choice, a constant star forming model 
will provide a poor fit. The dust model was originally calibrated to the same $m^*$ value as the others, but was not re-scaled to show the effect.}\label{fig:evol}
\end{figure*}


In the context of previous works, our results for the redshift dependence of $m^*$ are puzzling. \citet{man10} and \citet{bro13} 
analyzed the \Spitzer LFs and SFR (respectively) of the same cluster sample at $z<2$. They conclude 
that the epoch at which cluster galaxies are undergoing rapid mergers (and therefore mass assembly) is 
approximately $z\sim1.5$. \citet{wyl14} investigated the \Spitzer LF of clusters and protoclusters at $z\le3$,
and find no evidence of such a rapid mass assembly at $z\sim1.5$. They find that a passively evolving stellar
population is consistent with the full $m^*(z)$ range of both \citet{man10} and \citet{wyl14}.
\citet{wyl14} hypothesize
that as they are probing higher redshifts, and thus rarer/denser volumes, they do not
observe the mass assembly seen at $z\sim1.5$ by \citet{man10}, which correspond to more common 
overdensities. As a result, they speculate that it might be possible to observe a different 
epoch of cluster assembly at higher redshifts ($z>3$) than their sample. This
is akin to a cluster-scale version of galaxy downsizing, where the densest clusters will form the quickest. We do not see 
any behavior analogous to the high redshift ($1.5<z<2.0$) $m^*$ variability of \citet{man10} in our sample, which would indicate rapid mass assembly.

When our data is analyzed with the two previous works (Fig~\ref{fig:evol}), the evolution of the characteristic magnitude ($m^*$)
is fully consistent with a simple stellar population formed at $z_f\le10$. Even more puzzling is that the field sample 
at these redshifts is equivalent in all respects to the overdense sample, apart from the scaling. We gain little insight
into the epoch of rapid mass assembly of cluster galaxies, or of the field for that matter. More complex stellar populations, 
such as an exponential decay model with $\tau\sim1$ Gyr, can also be consistent with the data. A constant star formation model 
is not favored, regardless of which $m^*(z)$ value it is calibrated to.

It should be noted that this is not a progenitor-matched study, in that as the redshift increases, the possibility
of detecting a weak overdensity will decrease. CCPC candidates at $z>4$ will likely not evolve into the $z=2$ candidates
from \citet{wyl14} in a one-to-one fashion. However, some overlap could occur, particularly with their strongest 
overdensities. From our simple analysis, it appears that \Spitzer $m^*$ values in dense environments, a tracer of the stellar mass content of these galaxies,
is consistent with the passive evolution of a single burst of star formation at $z_f=10$ over nearly 10 Gyrs ($0.3<z<6.6$) of time.

It could be hypothesized that the marginally overdense CCPC candidates could merely be 
field galaxies. Their inclusion in the luminosity functions (Fig~\ref{fig:CH1_TOT_FIELD} and Fig~\ref{fig:CH2_TOT_FIELD})
could thus mask a weak signal of differentiation in the stellar mass functions of protocluster versus field galaxies. As a 
simple test of this hypothesis, we re-computed the [4.5] $m^*$ values for sub-samples of galaxies that exist in overdensities 
of $\delta_{gal} > 2,3,4$ and $5$. The redshift and magnitude bins remained unchanged from the analysis of the total sample.
The $m^*$ redshift evolution of these subsamples are broadly consistent with the $2\sigma$ uncertainties presented Table~\ref{tab:ch2_LF}.
There appears to be no statistically significant correlation between $m^*$ and $\delta_{gal}$ within our limited
sub-samples. For CCPC galaxies at $z<3.25$, $m^*$ is brighter by less than 0.3 mags between the subsamples of $\delta_{gal}> 4$ 
and $\delta_{gal}> 2$, on average. This is consistent with the uncertainties listed in Table~\ref{tab:ch2_LF}.
Significant deviations between the total and sub-samples of $\Delta m^* > 1$ occur only in some high redshift, 
high overdensity bins that lack significant galaxy numbers ($N\sim10^1$) to provide a satisfactory solution.


\subsection{Galaxy Selection Implications}

A further curiosity is evident when the evolution of $m^*$ is considered in the context of the galaxy type predominantly represented in our sample. CCPC galaxies
generally originate from LBG and line-emitting galaxy ($Ly\alpha$ emitters) spectroscopic surveys, which are selected directly because of their large UV-luminosity/SFR. Fig~\ref{fig:evol} 
illustrates that a consistent fit to the model of $m^*$ galaxies from $0.3<z<6.6$ is a simple stellar population, not a constant star formation model, dusty
or otherwise. At low redshifts this is not surprising, as a passive stellar population model has historically been well fit to overdense regions \citep{sta95}. However,
the high redshift systems are star forming galaxies \emph{by selection}, and yet the data clearly disfavor the model. The exponential decaying SFR model \citep{tin72} with $\tau=1$ Gyr
is also consistent with the evolution of $m^*$. However, this type of model is not able to fit roughly $1/4$ of all $z=0$ galaxy SFHs \citep{oem13}.
Exponential decay models with $\tau=10$ Gyrs look similar to that of the constant star formation model.
Fig~\ref{fig:evol} is normalized to match the data of \citet{man10} at redshift $z<1.1$. 

It could be assumed that by normalizing
the star formation model to a different $m^*(z)$ value, the data might be better fit. However, the shape of the model does not change,
just its scaling. The constant SFR model will be too bright at lower redshifts and too faint at higher redshifts, regardless of the scaling.
However, this simple observation should not be considered wholly unreasonable. There were a variety of individual sources that were fit with SED templates of exponential decaying SFRs, which 
are still forming stars at these early epochs of the universe, and the $\tau=1$ Gyr model in Fig~\ref{fig:evol} is consistent with the data. At these 
wavelengths, the predictions between a young, SSP and a decaying SFR cannot be disentangled within the uncertainty of the data.

An important factor to consider in the context of this entire work (not just the $m^*$ evolution) is that our spectroscopic galaxy sample does not
consist of the majority of galaxies at $z>2$. In fact, \citet{van06} showed that approximately $80\%$ of galaxies are not LBGs between $2<z_{phot}<3$. Many 
of these other objects are distant red and dusty star-forming galaxies (DRGs and DSFGs). This trend appears to become even more pronounced at higher redshifts.
In the range of $ 3<z<4$, only $14\%$ of galaxies with photometric redshifts would be identified as LBGs, with the remainder being nearly split between DSFGs and DRGs \citep{spi14}. 
It is unclear, without spectroscopic confirmation of protocluster membership, how these DRGs/DSFGs might cluster differently than their LBG counterparts,
or if they might have LFs that vary with environment. 

These DRGs are forming a not-insignificant amount of stars, on the order of a 20$\%$ contribution to the cosmic star formation density
at $1.5<z<2.5$ \citep{web06}. They therefore may not be completely unlike the spectroscopic galaxies in our sample. Unfortunately, \citet{spi14} noted that most of these DRGs are much fainter ($\sim2$ mags) 
than the canonical spectroscopic limit of $R_{AB}\le25.5$ for current instrumentation. For the present, it appears that this question will remain unanswered 
in the context of a spectroscopically-confirmed protocluster sample like the CCPC.

Although we do not expect to have DRGs/DSFGs in the CCPC, which are the dominant galaxy populations at high redshift, we can make comparisons
to cluster and protocluster candidates that do have these systems at lower redshift. In fact, we have already performed such an analysis, in that \citet{man10}
and \citet{wyl14} do not rely solely on spectroscopic redshifts for cluster membership. \citet{man10} compute photometric redshift probabilities
from deep, multi-wavelength data, while \citet{wyl14} utilize \Spitzer color cuts to identify high redshift galaxies in their 
overdensities. Both of these techniques are sensitive to galaxy populations not characterized by bright UV continuum
selection (e.g. LBGs). As the entirety of the $m^*$ evolution is consistent with a simple stellar population, passively evolving, 
it is plausible that the LFs of DRGs and spectroscopically confirmed LBGs may not be significantly divergent at these 
redshifts. Indeed, the redshifts at which \citet{wyl14} and this work overlap are in
agreement within the uncertainties, despite the differences in selection.


\subsection{Inferred Stellar Masses and Galaxy Properties}

SED fitting is an incredibly useful tool for estimating redshifts and 
galaxy properties at high redshift over a range of populations \citep{hyperz,van06}. However, it is 
possible to fit more models and parameters to the data than can actually be
constrained. In addition, there are significant degeneracies among model parameters that can match 
the same data at these high redshifts \citep{pap01,mag10}, such as the well known age-metallicity-dust degeneracy. The models differ
among themselves, with varying treatment of thermally-pulsating asymptotic giant branch stars 
or the adoption of varying IMFs \citep{bc03,basti}. Furthermore, with little \emph{a priori}
knowledge of the uncertain galaxy zoo extant at high redshift, the difficulties compound.
These issues are indeed true in the case of the CCPC, with our limited wavelength
coverage, as well as in numerous other studies. However, some properties can be loosely constrained
by our data, and is therefore a useful exercise if one is cognizant of the limitations of SED fitting.

Primarily, the rest-frame NIR data provided by \Spitzer provides a proxy for the underlying stellar
mass of the CCPC galaxies at high redshift. Unfortunately, for the same reason they are a powerful tool
for measuring stellar mass, these colors provide little information in determining 
any other property of the underlying galaxy \cite[e.g. passive versus star forming, metallicity variations,
formation redshift; see][]{coo14}. \HST filters (e.g. F125W, F140W, F160W) that measure rest-frame optical bands at the 
redshifts of the CCPC are able to generally distinguish between passive and star-forming galaxies.
In Fig~\ref{fig:HST_COL}, a number of model stellar populations are plotted as a function of redshift
and F160W-[4.5] color. A clear bifurcation is shown at $z\sim2$ which grows more pronounced at larger redshifts.
Within the photometric errors of our colors, it is not possible to assign a preferred star formation rate or formation
redshift to the CCPC galaxies. Clearly a range of stellar populations may exist. Dust obscuration is also uncertain, as a typical reddening of these colors is 
$\sim0.5$ mags or less for the BaSTI SSPs. This which is the $1\sigma$ photometric uncertainty in many cases.

To briefly illustrate the perils of  mass estimation among various models, let us take an idealized example of a galaxy with a measured $[4.5] = 20$ AB magnitude.
We can infer the stellar mass of this system from a menagerie of models available using a formation redshift $z_f=10$ and solar metallicity. At $z=2$,
the observed system can have a range of $1\times10^{11}<M_{\star}<3\times10^{11}$ $\Msun$  for the extreme cases of a BC03 SFG to a BaSTI SSP, respectively. At a redshift $z=6$, this 
gap can widen to roughly $4.5\times10^{11}<M_{\star}<9.5\times10^{11}$ $\Msun$. Lowering the formation redshift will also decrease the implied stellar mass by 
a factor of $\le3$. Notice that this example did not take into account any photometric uncertainties, dust, metallicity variations, or flux measurements at other wavelengths.
A change in $z_f$ can have an outsized role if it is close to the redshift of the galaxy (e.g. a young galaxy). An old, bright galaxy can have 10$\times$ 
the stellar mass than that of a young system of the same luminosity.

In practice, the $\chi^2$ values from successive SED fits were observed to not change significantly. This
was in spite of their sometimes drastically different stellar populations (e.g. quiescent versus star forming). A brief 
examination of Figure ~\ref{fig:HST_COL} reveals that various models can exist within the color uncertainty
of the CCPC galaxies. We wish to caution readers that SED fitting can have a difficult time \emph{excluding} models, especially with 
the limited wavelength coverage presented here. Therefore, stellar mass uncertainties are dominated by 
systematic variations in the models and the subsequent fitting procedure.  


Comparing the model fits on a system-by-system basis provides a cautionary tale
for determining galaxy properties (mass, age, metallicity) via the SED fitting method.
We applied our algorithm, for both log-normal and more traditional SEDs, to the 
data we measured, in addition to a companion photometric catalog in the CANDELS fields \cite[3D-HST;][]{ske14}.
The stellar mass estimates could vary by an order of magnitude or more for a single galaxy,
but we found \emph{no} statistically significant trends among the combination of two data sets and the two
SFH prescriptions (log-N versus traditional). Although there might appear to be a mean offset 
of formation redshift between our data and the 3D-HST catalog of $<z_f - z_f(3D)> \sim 2$,
for instance, the scatter between the two ($\sigma=4.5$) is much larger. The mean mass 
and $\chi^2$ differences follow much the same pattern, where occasionally a galaxy will be better fit or more massive
via log-normal fitting, but a subsequent galaxy will have the opposite effect. This appears to be a classic
case of overfitting the data, with various models supporting divergent implications (SFG vs. quiescent) being
equal fits to the photometry. It does not appear we are able to constrain the stellar populations
or masses for the CCPC galaxies with any reliability. 

Despite these concerns, our estimated stellar masses are not wholly unreasonable. We compare our 
mean $M_{\star}$ values to the sample investigated by \citet{mag10} of LBGs at $z\sim3$ with \Spitzer data.
Their mean stellar masses are $2.8\times10^{10}$ $\Msun$ and $4.2\times10^{10}$ $\Msun$, depending on 
which suite of models they use. Our mean values are  $3.0\times10^{10}$ $\Msun$ for log-normal fitting and
$6.7\times10^{10}$ $\Msun$ using our traditional SEDs. Their catalog also contains a few very bright, non-AGN 
sources that exceed $M_{\star}\ge5\times10^{11}$ $\Msun$, much like our own results (the hyperluminous sources).
Although individual objects may suffer from systematic uncertainties in the $M_{\star}$ estimates,
the CCPC as a whole is a reasonable match to other stellar mass studies of bright galaxies at 
high redshift.

\section{Summary} \label{sec:sum}

Although longitudinal data is required to perfectly map the evolution of galaxies \citep{abr16}, astronomy 
must content itself with studies that contain as minimal inherent bias as possible. This manuscript 
details the \Spitzer photometry of protocluster galaxies in the Candidate Cluster and 
Protocluster Catalog. The catalog probes galaxies between redshifts $2<z<6.6$ in dense environments. We built 
luminosity functions of the galaxies in various redshift bins at 3.6 and 4.5 $\mu$m wavelengths. 
These measure the rest-frame NIR emission of the galaxy populations to trace their stellar mass
as a function of redshift. 

The galaxies in both the field and CCPC samples contain extremely bright sources up to
5 magnitudes brighter than the characteristic magnitude $m^*$. These galaxies are divergent from 
the shape of the Schechter function, and exist at nearly all redshifts. Many of these are expected to be
bright AGN and a few ($<10\%$) low redshift interlopers. Semi-analytic models do not predict that these types of 
galaxies should exist \citep{guo11}, although they have been observed previously at similar 
redshifts \citep{per08,mar09}. Their nature is not yet established.

Field samples of galaxies are also photometrically measured, and remarkably the luminosity functions of the 
overdense regions are not statistically distinct from their field counterparts. In our current
understanding of galaxy formation, the expectation is that the most massive galaxies at any epoch
will be found in the densest environments. In Section ~\ref{sec:dis} we analyzed a number of possibilities 
that might explain this phenomenon, but cannot find a satisfactory conclusion. We believe this
to be the most fundamental result of this work.

We model the fitted LF parameter $m^*$ as a function of redshift in the context of various
stellar population models. By including the measurements at lower redshifts from \citet{man10}
and \citet{wyl14}, we find that a passively evolving stellar population formed in a single 
burst at high redshift ($z_f=7-10$) is consistent with the data at all redshifts ($0.3<z<6.6$). An exponentially decaying 
star formation model with $\tau=1$ Gyr is also in agreement with the data. Despite the fact that the majority 
of CCPC galaxies were spectroscopically selected based on their star forming properties (e.g. LBGs and line emitters)
a constant star forming model is a poor fit to the observed $m^*(z)$.

A SED fitting technique has provided stellar mass estimates and some general
information about the properties of the CCPC galaxies. We use supplemental \HST
data to probe the rest-frame optical emission to measure stellar colors for additional
model constraints. However, we are careful to note that even with greater wavelength coverage
than that which is presented here, SED fitting can be fraught with degeneracies (dust, age, metallicity) 
and inter-model uncertainties. Overall, the CCPC appears to be composed of $M\gtrsim10^{10}$ $\Msun$ 
galaxies with mean formation redshifts $z_f>7$. 
Apart from these broad statements, we cannot provide reliable dust content, metallicity 
information, or unique model fits to individual CCPC sources.

We thank the anonymous referee for helpful comments that improved 
the quality of this work. This work is based [in 
part] on observations made with the Spitzer Space Telescope, which is operated by the
Jet Propulsion Laboratory, California Institute of Technology under a contract with NASA.
This work has made use of the Rainbow Cosmological Surveys Database, which is operated 
by the Universidad Complutense de Madrid (UCM), partnered with the University of California 
Observatories at Santa Cruz (UCO/Lick,UCSC). This research has made use of the 
NASA/IPAC Extragalactic Database (NED) which is operated by the Jet Propulsion 
Laboratory, California Institute of Technology, under contract with the National 
Aeronautics and Space Administration. 
Based on observations made with the NASA/ESA Hubble Space Telescope, and obtained from the Hubble Legacy Archive, which is a collaboration between the Space Telescope Science Institute (STScI/NASA), the Space Telescope European Coordinating Facility (ST-ECF/ESA) and the Canadian Astronomy Data Centre (CADC/NRC/CSA).
This work is based on observations taken by the CANDELS Multi-Cycle Treasury Program with the NASA/ESA HST, which is operated by the Association of Universities for Research in Astronomy, Inc., under NASA contract NAS5-26555.

\bibliography{protos_spitzer}
\bibliographystyle{apj}
\appendix

The spectroscopic redshifts used in the Field samples of galaxies (Section ~\ref{sec:results}) came from the following sources (and references therein), as compiled by NED:
\citet{2008AnA...487..539W,2004AJ....128..544M,2004AnA...418..885N,2005AnA...439..845L}
\citet{2012ApJ...753...95B,2010AnA...512A..12B,2008ApJ...682..985W,2009AnA...504..751S,2004AnA...428.1043L}
\citet{2012ApJS..203...15B,2008ApJS..179...19L,2008ApJS..179...95M,2011ApJ...729...48B,2011AJ....141...14S}
\citet{2005AnA...437..883M,2010ApJ...720..368X,2011ApJ...743..144T,2009ApJ...693.1713T,2010MNRAS.405.2302H,2009ApJ...706..885W}
\citet{2013ApJ...772..113T,2003MNRAS.344..169E,2002AJ....124.1839B,2009ApJ...699..667M,2006ApJ...653.1004R}
\citet{2006MNRAS.370.1185P,2011MNRAS.413...80C,2013ApJ...772...48P,2006ApJ...646..107E,2004ApJ...604..534S,2001ApJ...559..620P}
\citet{1999ApJ...513...34F,2009ApJ...705...68B,2001ApJS..135...41F,2009ApJ...691..140H,2009ApJ...697.1410L}
\citet{2009AJ....137..179R,2011ApJ...735...86W,2010ApJ...717.1181P,2012ApJ...759L..44G,2013ApJ...776....9G,gob11}
\citet{2009ApJ...690..295K,2003ApJ...591..101E,2005ApJ...626..698S,2006ApJ...648..250C}
\citet{2002AJ....123.3041K,2003ApJ...592..728S,2004AJ....127.2455A,2002AnA...396..847V,2007ApJ...655...51W,2005ApJ...633.1126K}
\citet{1998ApJ...499L.135S,2013AnA...559A...2G,2006AnA...457...79G,2006AnA...450..495W,2010ApJS..191..124S}
\citet{2004ApJS..155..271S,2010ApJ...722.1895E,2013ApJ...765L...2B,2014ApJ...793..101T,2008ApJ...677..219K,1999AnA...343..399W}
\citet{2010AnA...509A..83D,2010ApJ...716..348B,2012MNRAS.423.2436S,2014ApJ...791....3S}
\citet{2002AnA...396..109P,2004AnA...428..817K,2000AnA...361L..25P,2005AJ....130..867C,2004AnA...428..793K,1997AnA...326..505R}
\citet{1997ApJ...481..673L,2005ApJ...631..101P,2002AJ....124.1886M,2004ApJ...613..655W,2011PASJ...63S.437T}
\citet{2010ApJ...718..112Y,2011MNRAS.416.2041M,2005ApJ...620..595W,2006MNRAS.371..221G,2004ApJ...616...71S,2010ApJ...715..385O}
\citet{2004ApJ...612..108S,1995AnA...294..377V,2007ApJ...668...23P,2004ApJ...614..671C,2000AnA...362....9M}
\citet{2000AJ....120.1648C,2010MNRAS.401..294S,2013AnA...557A..81M,2014ApJ...788..125S,2010ApJ...719.1393D,2011MNRAS.411.2739C}
\citet{2009AnA...501..865S,2006AnA...449..951G,2014MNRAS.440.3630R,1994ApJ...436..678O}
\citet{2005ApJ...618..123S,1996ApJ...468..121S,2004ApJ...606..664D,2004AJ....127.3137C,2004ApJ...612..122E,2011ApJS..192....5A}
\citet{2000AJ....119.2092B,2009ApJ...703..198L,1997ApJ...489..543P,2008ApJS..179....1T,2004ApJ...617...64S}
\citet{2004AnA...424....1D,1987ApJ...314..111A,2010MNRAS.407..846D,2011ApJ...740L..31E,2012MNRAS.426.1073H}
\citet{2009MNRAS.400..299L,1999AnA...345...73P,1997ApJ...478...87H,1991MNRAS.250...24Y,1995AJ....109.1522C}
\citet{2004ApJS..155...73Z,2014AJ....148...13R,2011ApJ...735...87Y,2007ApJ...660..167D,2011AnA...526A..86G,2004AJ....127.3121W}
\citet{2006ApJ...637L...5D,2012ApJ...745...33K,2012ApJ...745...85L,1996ApJ...457..102D,1997ApJS..112....1T}
\citet{1996AJ....112...62T,2011ApJ...733...31H,2001ApJ...560..127S,2012ApJ...759..139K,2009AnA...507.1277B,2014ApJ...791...18B}
\citet{2005ApJ...626..680D,2005MNRAS.359..895V,2008MNRAS.384.1611K,2002AJ....123.1163B,2001AJ....122..598D}
\citet{1998AJ....115.1400F,2003AJ....126.1183C,2011ApJ...733L..11R,2011MNRAS.412.1913I,2006ApJ...647...74W,2001AJ....121..662B}
\citet{2001AJ....122.1125P,1989AnA...218...71R,2005AnA...440..881I,2012AnA...537A..31C,2011ApJ...736...48R}
\citet{2007ApJ...659..941T,die13,2013MNRAS.429.3047B,1996Natur.381..759L,1999ApJ...511L...1C,2008MNRAS.389...45C}
\citet{2000ApJ...545..591W,2013MNRAS.430..425B,2009ApJ...703.2033R,2008ApJ...675..262R,2005ApJ...629...72A}
\citet{2012ApJS..199....3R,2007ApJS..172...70L,2002MNRAS.337.1153S,2001AJ....122.2177B,1995ApJ...448..575R,2006ApJ...637..648S}
\citet{ven07,1996ApJS..107...19M,2007AnA...467...63T,2004ApJ...606..683S,2001AnA...380..409V}
\citet{1998AJ....115.2184S,2011ApJ...731...97S,2004ApJ...611..732C,mag10,2003AnA...407..147F,2000ApJ...543..552S}
\citet{2008AnA...492..637G,2000AnA...359..489C,2011ApJ...729L...4F,2001ApJ...562...95S,1995MNRAS.277..389R}
\citet{2011ApJ...741...91C,2001AnA...374..443F,2001AnA...372L..57M,2004AJ....127..131S,2003ApJ...597..680W,1997AnA...318..347S}
\citet{1996ApJ...462...68S,2005AnA...431..793V,1996ApJ...471L..11L,2008MNRAS.389.1223M,2012ApJ...744..110C}
\citet{2010ApJ...716L.200B,2008AJ....135.1624S,2007AJ....134..169X,2009ApJ...691..687L,2007ApJ...657..135C,1998AJ....115...55L}
\citet{2009ApJ...696.1195T,2009AnA...497..689G,2008ApJ...689..687B,1998ApJ...494...60P,2005ApJ...619..134T}
\citet{2005ApJ...627...32S,1993MNRAS.260..202H,1995AnA...296..665V,2011MNRAS.411.2336I,2011ApJ...736...18N,2006ApJ...651..688S}
\citet{2004ApJ...606...85C,2009MNRAS.396L...1W,2004AJ....128.2073H,2000ApJ...532..170S,2004AnA...423..761V}
\citet{1999ApJ...519....1S,2002AnA...393..809M,2001ApJ...549..770E,1996AnA...305..450d,2008AnA...492...81P,2010MNRAS.401.1657L}
\citet{2008AnA...483..415K,2012AnA...537A..16F,2006AnA...454..423V,2014ApJ...793...92T,2011AnA...528A..88G}
\citet{2005AJ....130..496J,2011ApJ...736...41B,2001ApJ...554..742H,2003AJ....126..632B,2006ApJ...650..614H}
\citet{2000ApJS..130...37S,2000MNRAS.318..817S,2008ApJS..176..301O,2008AnA...478...83V}
\citet{2007AnA...461...39F,2009ApJ...698..740T,2009ApJ...693....8B,1992MNRAS.259p...1P,2008ApJ...675.1076S}
\citet{2011ApJ...738...69S,2010ApJ...725.1011V,2009ApJ...697..942R,2009ApJ...695L.176D,2004ApJ...611..660O}
\citet{2006ApJ...653..988Y,1994AnA...281..331S,1991AJ....101.2004S,2007MNRAS.376.1557I,2006AnA...455..145T,2009ApJ...706..762W}
\citet{2004ApJ...617..707D,2006MNRAS.372..357M,2005ApJ...620L...1O,2006ApJ...648...54S,2009AnA...500L...1M}
\citet{2010ApJ...725..394H,2013AnA...555A..42R,2013MNRAS.431.3589Z,2009MNRAS.393.1174F,2009ApJ...691..465F,2010AnA...510A.109R}
\citet{2007MNRAS.376..727S,2005AnA...434...53V,2010MNRAS.407L.103C,2011ApJ...743..132P,2004MNRAS.355..374B}
\citet{2005ApJ...626..666M,2005ApJ...621..582R,2014AnA...564A.125B,2008ApJ...673..686H,2007MNRAS.376.1861F,2004ApJ...604L..13S}
\citet{2007ApJ...660...47D,2011ApJ...735....5F,2013MNRAS.432.2869H}
\citet{2001AJ....121.1799P,2003ApJ...596...67D,2012ApJ...761..139C,2009ApJ...701..915T,2008ApJ...679..942M,2012ApJ...744..149H}
\citet{2007ApJS..172..523M,2010MNRAS.408L..31D,1996ApJ...459L..53H,2013ApJ...763..120C,1996ApJ...456L..13E}
\citet{1996AnA...316...33W,2006ApJ...650..604R,2009ApJ...704..117K,2013ApJ...763..129S,2011ApJ...728L...2S,2007ApJ...668..853K}
\citet{1999PASP..111.1475S,1998ApJ...505L..95W,2002ApJ...570...92D,2014AnA...562A..35N,2002AstL...28....1R}
\citet{1998AJ....116.2617S,2004ApJ...610..635A,2014AnA...569A..98T,2011ApJ...734..119K,2006PASJ...58..313S,2014ApJ...792...15T}
\citet{2013ApJ...772...99J,2007AnA...468..877N,2007ApJ...671.1227D,2009ApJ...700...20L,2001AJ....122..503A}
\citet{2004AJ....127..563H,1999ApJ...522L...9H,2012MNRAS.422.1425C,2010ApJ...723..869O,2012AnA...547A..51G,2009ApJ...696.1164O}
\citet{2012ApJ...744..179S,2012ApJ...744...83O,2004ApJ...607..704S,2011ApJ...743...65J,2005PASJ...57..165T}
\citet{2006ApJ...648....7K,2006Natur.443..186I,2004ApJ...613L...9N,2005ApJ...634..142N,2004ApJ...611...59R}

\end{document}